\documentclass[traditabstract,fleqn]{aa}
\usepackage{epstopdf}
\usepackage{natbib,amssymb,amsmath,txfonts,astro,url,mathtools}
\usepackage{longtable,booktabs}
 \usepackage{lscape}
\usepackage{diagbox}
\usepackage{cellspace}
\usepackage{multirow}
\usepackage{slashbox}
\DeclareGraphicsExtensions{.pdf,.png,.eps,.gif,.ps}
\begin{document}
\setlength{\mathindent}{0pt}
\newcommand{\dg}{$^\circ$ }
\newcommand{\kmss}{\,km\,s$^{-1}$}
\newcommand{\cmt}{cm$^{-3}$}
\newcommand{\cmtwo}{cm$^{-2}$}
\newcommand{\cya}{HC$_3$N}
\newcommand{\meCN}{CH$_3$CN}
\newcommand{\acetA}{CH$_3$CHO}
\newcommand{\forma}{NH$_2$CHO}
\newcommand{\cmC}{cm$^{-3}$}
\newcommand{\cycloprop}{{\it c-}C$_3$H$_2$}
\newcommand{\Ncycloprop}{$N_{c\text{-C}_3\text{H}_2}$}
\newcommand{\NHCO}{$N_{\text{HCO}^+}$}
\newcommand{\cyclopropIso}{{\it c-}H$^{13}$CCCH}
\newcommand{\linprop}{{\it l-}C$_3$H$_2$}
\newcommand{\meOH}{CH$_3$OH}
\newcommand{\meOHiso}{$^{13}$CH$_3$OH}
\newcommand{\meOHIsoRat}{CH$_3$OH/$^{13}$CH$_3$OH}
\newcommand{\oformald}{{\it o-}H$_2$CO}
\newcommand{\pformald}{{\it p-}H$_2$CO}
\newcommand{\formald}{H$_2$CO}
\newcommand{\formaldIso}{H$_2^{13}$CO}
\newcommand{\HCS}{HCS$^{+}$}
\newcommand{\HCO}{HCO$^{+}$}
\newcommand{\HtwoCS}{H$_2$CS}
\newcommand{\CtwoH}{C$_2$H}
\newcommand{\cisC}{{\it cis-}CH$_2$OHCHO}
\newcommand{\delv}{$\Delta$v}
\newcommand{\gsim}{$\gtrsim$}
\newcommand{\lsim}{$\lesssim$}
\newcommand{\AV}{$A_{\text V}$}
\newcommand{\AVo}{$A_{\text{V}o}$}
\newcommand{\TLTC}{$T_L$/$T_C$~}
\newcommand{\TL}{$T_L$~}
\newcommand{\TC}{$T_C$~}
\newcommand{\TCMB}{$T_\text{CMB}$}
\newcommand{\Tex}{$T_{ex}$~}
\newcommand{\EU}{$E_U$~}
\newcommand{\EL}{$E_L$~}
\newcommand{\lBrett}{{\it l-}C$_3$H$^+$}
\newcommand{\lneutralBrett}{{\it l-}C$_3$H}
\newcommand{\zero}{$\sim$0~}
\newcommand{\lineClouds}{-106, -73, -40, and 0~\kms}
\newcommand{\SijU}{$S_{ij}\mu^2$}
\newcommand{\LOS}{line-of-sight}
\newcommand{\los}{line-of-sight}
\newcommand{\LOScap}{Line-of-sight}
\newcommand{\StoN}{signal-to-noise}
\newcommand{\twelveThirteen}{$^{12}$C/$^{13}$C}
\newcommand{\twelveCOThirteenCO}{$^{12}$CO/$^{13}$CO}
\newcommand{\twelveHCOThirteenHCO}{H$^{12}$CO$^{+}$/H$^{13}$CO$^{+}$}
\newcommand{\thirtyTwoThirtyFour}{$^{32}$S/$^{34}$S}
\newcommand{\twentyEightTwentyNine}{$^{28}$Si/$^{29}$Si}
\newcommand{\twentyEightThirty}{$^{28}$Si/$^{30}$Si}
\newcommand{\twentyNineThirty}{$^{29}$Si/$^{30}$Si}
\newcommand{\GC}{Galactic Center}
\newcommand{\Nmol}{$N_{molec}$}
\newcommand{\CSiso}{C$^{34}$S}
\newcommand{\Htwo}{H$_2$}
\newcommand{\NHtwo}{$N_{\text{H}_2}$}
\newcommand{\NHI}{$N_{\text{H{\sc i}}}$}
\newcommand{\XCS}{$X_{CS}$}
\newcommand{\XmeOH}{$X_{\text{CH}_3\text{OH}}$}
\newcommand{\twelveC}{$^{12}$C}
\newcommand{\elec}{$e^-$}
\newcommand{\Hthreeplus}{H$_3^+$}
\newcommand{\sigrms}{$\sigma_{rms}$}
\newcommand{\HII}{H{\sc ii}~}
\newcommand{\Halpha}{H\,$\alpha$~}
\newcommand{\HI}{H{\sc i}~}
\newcommand{\OM}{order of magnitude}
\newcommand{\OMs}{orders of magnitude}
\newcommand{\NCS}{$N_{\text{CS}}$}
\newcommand{\NCCS}{$N_{\text{CCS}}$}
\newcommand{\NSO}{$N_{\text{SO}}$}
\newcommand{\NCO}{$N_{\text{CO}}$}
\newcommand{\NCH}{$N_{\text{CH}}$}
\newcommand{\NCCH}{$N_{\text{CCH}}$}
\newcommand{\NHCS}{$N_{\text{HCS}^+}$}
\newcommand{\Nformald}{$N_{\text{H}_2\text{CO}}$}
\newcommand{\NHtwoCS}{$N_{\text{H}_2\text{CS}}$}
\newcommand{\NOH}{$N_{\text{OH}}$}
\newcommand{\oneOsix}{-106}
\titlerunning {The Molecular Chemistry of Diffuse and Translucent Clouds in the Line-of-Sight to Sgr B2}
\title{The Molecular Chemistry of Diffuse and Translucent Clouds in the Line-of-Sight to Sgr B2: Absorption by Simple Organic and Inorganic Molecules in
  the GBT PRIMOS Survey}
\authorrunning {J.~F.~Corby et al.}
\author{J.~F.~Corby \inst{1,2}, B.~A.~McGuire \inst{2}, E.~Herbst \inst{3}, A.~J.~Remijan \inst{2}}

\institute{$^1$ Department of Physics, University of South Florida, 4202 East Fowler Ave, Tampa, FL 33605, USA \\
 $^2$ National Radio Astronomy Observatory, 520 Edgemont Rd, Charlottesville, VA 22903, USA \\
 $^3$ Departments of Chemistry and Astronomy, University of Virginia,  McCormick Road, Charlottesville, VA 22904, USA}

\date{Received date: 18 April 2017; Accepted date 9 August 2017}
 
\abstract{
The 1-50 GHz PRebiotic Interstellar MOlecular Survey (PRIMOS) contains $\sim$50 molecular absorption lines observed in clouds located in the 
\los ~to Sgr B2(N). The \los~material is associated with diffuse and translucent clouds located in the Galactic Center, Bar, and spiral arms in the disk. 
We measure the column densities and estimate abundances, relative to \Htwo, of 11 molecules and additional isotopologues observed in this material. 
We use absorption by optically thin transitions of \cycloprop~to estimate the molecular hydrogen columns, and argue that this method is preferable to 
more commonly used methods.  We discuss the kinematic structure and abundance patterns of small molecules including the sulfur-bearing species
CS, SO, CCS, \HtwoCS, and \HCS; oxygen-bearing molecules OH, SiO, and \formald; 
and simple hydrocarbon molecules \cycloprop, \lneutralBrett, and \lBrett.    
Finally, we discuss the implications of the observed chemistry for the structure of the gas and dust in the ISM.  

Highlighted results include the following.  First, whereas gas in the disk has a molecular hydrogen fraction of 0.65, 
clouds on the outer edge of the Galactic Bar and in or near the Galactic Center have molecular fractions of 
0.85 and \textgreater0.9, respectively. 
Second, we observe trends in isotope ratios with Galactocentric distance; while carbon and silicon show enhancement of the 
rare isotopes at low Galactocentric distances, sulfur exhibits no trend with Galactocentric distance.  We also determine that
the ratio of \cycloprop/\cyclopropIso ~provides a good estimate of the \twelveThirteen ~ratio, whereas \formald/\formaldIso ~exhibits fractionation.
Third, we report the presence of \lBrett ~in diffuse clouds for the first time.
Finally, we suggest that CS has an enhanced abundance within higher density clumps of material in the disk, 
and therefore may be diagnostic of cloud conditions.  If this holds, the diffuse clouds 
in the Galactic disk contain multiple embedded hyperdensities in a clumpy structure, 
and the density profile is not a simple function of \AV.  
}

\maketitle

\section{Introduction}
\label{sec:Intro}

The diffuse ISM serves as a significant reservoir for gas in the Galaxy, 
and is believed to be involved in processes of gas accretion,
recycling and feedback from star formation, and in the first stages of molecular cloud formation.  
Thus, an understanding of this material is essential for understanding the processing of gas in galaxies.  
Although the first interstellar molecular detections were made in diffuse gas, namely CH, CH$^+$, and CN \citep{SwingsRosenfeld1937,McKellar1940,DouglasCHplus}, 
the diffuse ISM was not expected to have a very rich molecular chemistry, and recent observations have proven surprising.

While our picture of the physical and chemical structure of the diffuse ISM continues to evolve, the chemistry reveals three primary phases with blurred boundaries.  
First, the most diffuse material produces absorption primarily by certain light hydride species \citep[e.g.][]{Gerin2010, Qin2010, Indriolo2015}.  
The absorption components tend to be broad \delv ~\gsim~15 \kmss, indicating large spatial extents, and estimated physical conditions include low densities (50~\textless~$n$~\textless~300~\cmt), 
warm kinetic temperatures (80~\textless~$T$~\textless~300 K) \citep{Snow&McCall}, and low molecular hydrogen fractions $\left(f(\text{H}_2) = \frac{2 N_{\text{H}_2}}{N_{\text{H{\sc i}}} + 2 N_{\text{H}_2}} < 0.1\right)$ \citep{Indriolo2015, Snow&McCall}.
In the second phase, heavier molecules appear, including HCN, CS, \formald, and \cycloprop, 
but estimated physical conditions still include low densities (100~\textless~$n$~\textless~500~\cmt) and warm temperatures \citep{Snow&McCall, LL06,LisztcC3H2}.  
In this phase, higher molecular fractions are present, with $f(\text{H}_2) \approx 0.4$ \citep{LisztcC3H2}, and 
extinction conditions of \AVo~\gsim~0.2 mag (implying a total extinction \AV~\gsim~0.4) are typical \citep{Snow&McCall}.  

The third phase is a translucent cloud, which is historically defined as material 
with a central extinction of 1~\textless~\AVo~\textless~2.5 and more recently defined 
as a cloud where $f(\text{C}^+)$ \textless~0.5 and $f(\text{CO})$ \textless~0.9, so that carbon is present in
neutral, ionized, and molecular forms concurrently \citep{Snow&McCall}.  
Translucent clouds are embedded within diffuse clouds, and their kinematic signatures include 
absorption over the same general velocity range as the diffuse material, but with narrower features.  
Galactic disk translucent clouds have densities of order 5000~\cmt ~and cool temperatures of 15 \textless~$T$\textless~30 K.
In Galactic Center translucent clouds, densities may be higher, up to 10$^{4}$~\cmt~\citep{Greaves92,Greaves94}, 
and temperatures are much warmer at 50 \textless~$T$\textless~90 K.

In all three phases, standard chemical models underpredict the molecular abundances of most observed molecules by multiple \OMs~\citep{LL_CS,Godard12}, 
indicating a poor understanding of the physics and chemistry of these regions.  
Models incorporating turbulent dissipation, where material is heated and compressed periodically, produce model predictions that are more consistent with observations for light hydride species
in diffuse clouds \citep{Godard09,Godard12}.  Yet much work remains in order to understand the coupled physics and chemistry within the diffuse and translucent media, including 
the physical and chemical exchange between the three phases.

Furthermore, the diffuse and translucent gas in the Galactic Center appears to be systematically different from that observed in the disk by multiple measures. 
In addition to the temperature and density differences noted above, observations reveal very high abundances of \Hthreeplus ~with a filling factor of nearly 1 \citep{Oka05}
and a larger fractional volume filled by translucent clouds in the Galactic Center compared to the disk, due largely to the extreme amount of gas mass in the Galactic Center \citep{MorrisSerabyn96}. 
Cosmic ray (CR) ionization rates in diffuse gas in the Galactic Center have been measured to be enhanced by 1 to 3 \OMs ~compared to the disk \citep{lePetitCRio,Indriolo2015}, 
and significantly higher X-ray fluxes are also present \citep{AmoB09}.  
It is not precisely known how this affects the chemistry in these regions; several small molecules including \Hthreeplus, OH$^+$, and H$_2$O$^+$ are predicted 
and observed to have enhanced abundances in high CR flux environments, but for most larger neutral species, 
chemical models predict lower abundances due to molecular dissociation \citep[e.g.][]{Donghui_2016}.  

In this paper, we present absorption profiles by eleven molecules and five isotopologues in the \los ~to Sgr B2(N), as measured by the GBT PRebiotic MOlecular Survey (PRIMOS).
PRIMOS is the most sensitive broadband survey at centimeter wavelengths, conducted towards Sgr B2(N), a high-mass star forming region in the Galactic Center known for its extreme 
mass and diversity of molecular chemistry.  In the 1-50 GHz spectrum obtained by PRIMOS, foreground absorption is detected from $\sim$50 molecular 
transitions of 17 molecules and six isotopologues.  In this work, we present the kinematic structure 
and abundance patterns of the sulfur-bearing species
CS, SO, CCS, \HtwoCS, and \HCS; oxygen-bearing molecules OH, SiO, and \formald; 
simple hydrocarbon molecules \cycloprop, \lneutralBrett, and \lBrett; and detected isotopologues of these species.   
We exclude from this paper detected line-of-sight absorption by NH$_3$, as absorption in the metastable NH$_3$ transitions should be 
treated in a different analysis, and by complex organic molecules (COMs), which will be presented in a second paper (Paper II; Corby et al. in prep)
to closely follow.

While absorption and emission associated with Sgr B2 spans the velocity range of +40 to +90 \kmss, 
diffuse and translucent clouds believed to be located in the Galactic Center, Bar, and spiral arms are observed in the range of -130 to +40 \kmss.  
While the exact locations of most of the cloud components are not confidently known, the following guidelines are helpful for interpreting abundance patterns.  
Features in the velocity range of -130 to -55 \kms ~are believed to be located within 1 kpc of the Galactic Center, and although some of the material at 0 \kms ~is presumably local to the Orion Arm, 
most of the absorption from -10 to +25 \kms ~also originates from within or near the Galactic Center.  
Amongst the most prominent features, the -106 \kms ~cloud is associated with the Expanding Molecular Ring (EMR) that spans the Galactic Center and marks the outer edge of the 
CMZ at a 200 to 300 pc galactocentric distance \citep{WhiteoakGardner79}; 
the -73 \kms ~cloud is suggested to be part of the 1-kpc disk located deep within the Galactic Bar \citep{WirstromNH3}; and while some researchers have suggested that the \zero \kms ~component
consists of ejecta from Sgr B2 \citep{WirstromNH3}, multiple observations have shown that this feature is widespread across the Galactic Center, 
and is a rather unique system containing molecular line absorption and weak recombination line emission across the Galactic Center \citep{Jones12, MarcCar}, 
as well as weak masing in the viscinity of Sgr B2(N) \citep{Corby15}.  The gas from -55 to -35 \kms ~is associated with gas in the near 3-kpc arm located 
on the outer edge of the Galactic Bar, while the material from -35 to -10 \kms ~is believed to arise in a spiral arm in the disk at a Galactocentric distance of $\sim$4 kpc \citep{WhiteoakGardner79, WirstromNH3}.

The remainder of this paper is organized as follows.  In \S \ref{sec:obs}, we describe the observations.  
In \S \ref{sec:PRIMOSresults}, we present the line profiles (\S \ref{subsec:profiles}), 
cloud kinematics (\S \ref{subsec:kinematics}), and measured column densities and abundances 
(\S \ref{subsec:colDenssAbunds} and \S \ref{subsec:abundancePatterns}).  
In \S \ref{sec:Hydrogen}, we discuss the method for molecular hydrogen column estimation. 
In \S \ref{sec:discussion}, we discuss the results in the context of trends with Galactocentric distance and 
implications for the cloud structures, before concluding in \S \ref{sec:SACconclusions}.

\subsection{Molecular Excitation and the Line-to-Continuum Ratio}
\label{subsec:Tex}

The densities in diffuse and translucent clouds are insufficient to excite molecules much beyond the CMB temperature, 
and excitation temperatures within this material are observed to span the range of 2.7~K~\lsim~\Tex~\lsim 4~K  \citep{Greaves94, Linke81}.  
As a result, the equation for line optical depth reduces to $$\tau = -ln(1+\text{T}_L/{\text T}_C),$$ for baseline subtracted absorption strength \TL 
absorbing against a continuum source of strength T$_C$.  The value \TLTC fully characterizes the line optical depth, 
and the data are therefore presented in this scale.  
The scale additionally provides higher precision measurements, as many sources of calibration error that typically apply to 
single dish radio data cancel in this ratio, including atmospheric opacity corrections, telescope efficiencies, and absolute flux calibrations.


\section{Observations}
\label{sec:obs}

The observations were conducted with the Robert C. Byrd Green Bank Telescope (GBT) through the PRebiotic Interstellar 
MOlecular Survey (PRIMOS)\footnote{PRIMOS is publicly available at http://www.cv.nrao.edu/PRIMOS/.}. 
PRIMOS is a key science project of the GBT that provides the deepest, most frequency-complete centimeter-wave spectral survey completed to date.  For all observations, the 
telescope was pointed towards the position of the Large Molecule Heimat (LMH) at $\alpha$ = $17^h47^m19\fs8$, $\delta$ = $-28\degr22\arcmin17\arcsec$. 
The LMH is a hot core source of 5 arcsec in spatial extent located on the SW edge of the Sgr B2(N) continuum structure. The GBT beam,
which varies from 13 arcmin at 1 GHz to 15 arcsec at 50 GHz, is sensitive to
absorption in the foreground of the free-free continuum structure of Sgr B2(N) as well as molecular line emission from the LMH over the full range of frequencies.
At very low frequencies (of \textless 8 GHz), the beam contains Sgr B2(M) as well.
Most of the data were collected over the full year of 2007, with some observations performed in 2002, 2005, and 2013. All observations were performed 
in position switching mode, with an {\sc off} position located 1 degree E in azimuth. Over most of the frequency range (of \textgreater 3 GHz),
four spectral windows of 200 MHz bandwidth and 8192 channels were observed simultaneously; 50 MHz windows were observed at frequencies of \textless3~GHz.
Most windows were observed for $\sim$10-15 hours, resulting in $T_A^*\sim$5 mK rms noise levels in most of the frequency range.

The absorption profiles are carefully baseline subtracted over the narrow frequency ranges corresponding to about -400 to +250 \kms 
~using a polynomial solution.  
For data with very stable baselines (typically Q-Band), we used a 1st order baseline.  For data with more highly varying baselines, 
we usually used a 4th order polynomial fit over a typical velocity range of -400 to +250 \kmss. 
The data were then normalized by the continuum, as the ratio of \TLTC is the fundamental value for characterizing the molecular line absorption from the diffuse and translucent 
clouds.  This ratio is independent of corrections to the atmospheric attenuation and the GBT aperture efficiency, and these corrections therefore were not applied.

\section{Results}
\label{sec:PRIMOSresults}

In observations toward Sgr B2(N), two absorption components 
are associated with Sgr B2 itself, at +64 and +82 \kms ~\citep{HuttemeisterNH3, Corby15}.
We assume that line absorption and emission in the range of +40 to +90 \kms ~is from Sgr B2 and
refrain from further discussing this material, instead focusing on the molecular composition of clouds 
observed in the velocity range of -126 to +24 \kmss, associated with material in the line of sight to Sgr B2.
Throughout this section, we assume that excitation conditions in all absorbing components are equivalent and characterized by 
\Tex = 3~K.  This assumption has been validated in the line of sight to Sgr B2(M) for
CS \citep{Greaves94}, and excitation temperatures of 2.73 to 3~K have been widely adopted for studies of diffuse and translucent 
clouds in line-of-sight absorption
\citep[e.g.][]{Nyman84,GN96, LL93, Qin2010, LisztcC3H2, Wiesemeyer2016}.  The assumption is further validated in Section \ref{subsec:colDenssAbunds}, in which 
we explore the line excitation of molecules sampled by multiple transitions in the PRIMOS data.  By assuming
equivalent excitation in all line-of-sight clouds, it is possible to directly compare column densities in different clouds 
from a single line profile.


\subsection{Spectral Profiles} 
 \label{subsec:profiles}


Table \ref{tab:transitions} lists transitions that were observed in line-of-sight absorption with sufficient 
signal-to-noise and baseline stability to enable a fair characterization.  For all listed transitions, we 
fit Gaussian components over the velocity range of -126 to +24 \kms ~using the {\sc fitgauss} function provided 
in the {\sc gbtidl} data reduction package and/or with a Gaussian fitting function constructed by the authors. 
We attempted to use as few Gaussian components as could account for the line profile to within the noise level
and we determined errors to the fits as the square root of the diagonal entries of the covariance matrix.  
Gaussian fits are overlaid on all spectral profiles shown, and individual Gaussian components are 
shown with a dashed profile. 
For transitions with resolvable hyperfine structure or A/E splitting including lines of \lneutralBrett 
~(J=$\frac{3}{2}$-$\frac{1}{2}$, $\Omega=\frac{1}{2}$), we fit the primary hyperfine component and assumed
that the satellite lines have the same line shape with the height set by the relative line strengths of the
satellite and primary features.  Selected line profiles are shown throughout this subsection, and the full 
catalog of line profiles listed in Table \ref{tab:transitions} are shown in Appendix A. 


\begin{table*}[!htb] 
\centering
\small
\caption{Summary of Molecular Lines Observed in Line-of-sight Absorption}
\label{tab:transitions}
\begin{tabular}{llcccc}
\hline
Molecule    & Transition & Frequency & E$_L$ & $\theta_B$ & Catalogue\\
            &            & (MHz)     & (K)   &   &  \\

\hline

OH			& J=$\frac{3}{2}$ $\Omega$=$\frac{3}{2}$ F=1$^+$-2$^-$	& 1612.2310(2) & 0.0026 & 7.8\arcmin & JPL\\  
OH			& J=$\frac{3}{2}$ $\Omega$=$\frac{3}{2}$ F=1$^+$-1$^-$	& 1665.4018(2) & 0.000 & 7.6\arcmin & JPL\\
OH			& J=$\frac{3}{2}$ $\Omega$=$\frac{3}{2}$ F=2$^+$-2$^-$	& 1667.3590(2) & 0.0026 & 7.6\arcmin & JPL\\
\cycloprop 		& 1$_{10}$ - 1$_{01}$ 	& 18343.143(2)	& 2.35	& 41.2\arcsec 	& CDMS \\
\cycloprop 		& 2$_{20}$ - 2$_{11}$ 	& 21587.4008(3) & 8.67 & 35.0\arcsec 	& JPL\\
\cycloprop 		& 2$_{11}$ - 2$_{02}$	& 46755.610(2)	& 6.43	& 16.2\arcsec	& CDMS \\
{\it c-}H$^{13}$CCCH & 1$_{10}$ - 1$_{01}$ 	& 18413.8248(7)	& 2.29 & 41.1\arcsec 	& CDMS \\

\formald 		& 1$_{10}$ - 1$_{11}$ 	& 4829.660(1)	&   15.16 & 2.6\arcmin &  JPL\\
\formald 		& 2$_{11}$ - 2$_{12}$ 	& 14488.479(1)	& 21.92 & 52.2\arcsec &  JPL\\ 	
\formald 		& 3$_{12}$ - 3$_{13}$ 	& 28974.805(10)	& 32.060 & 26.1\arcsec	& JPL\\ 
H$_2^{13}$CO		& 1$_{10}$ - 1$_{11}$	& 4593.0885(1)	&  15.126 & 2.7\arcsec	&JPL\\
H$_2^{13}$CO		& 2$_{11}$ - 2$_{12}$	& 13778.8041(2)	&  21.723 & 54.9 \arcsec & JPL \\

$^{28}$SiO 		& 1 - 0  		& 43423.76(5) 	& 0.000 & 17.4\arcsec  & CDMS \\
$^{29}$SiO 		& 1 - 0 		& 42879.949(3) 	& 0.000 & 17.6\arcsec 	& CDMS \\
$^{30}$SiO 		& 1 - 0 		& 42373.427(5) 	& 0.000 & 17.8\arcsec 	& CDMS \\

SO			& 1$_0$ - 0$_1$		& 30001.58(10)	& 0.000 & 25.2\arcsec	& CDMS \\

CS			& 1 - 0			& 48990.955(2)	& 0.000 & 15.4\arcsec	& CDMS \\ 
C$^{34}$S		& 1 - 0			& 48206.941(2)	& 0.000 & 15.7\arcsec	& CDMS \\

CCS			& N$_J$ = 1$_2$ - 2$_1$	& 22344.031(1) 	& 0.533	& 33.8\arcsec	& CDMS \\
CCS			& N$_J$ = 2$_3$ - 1$_2$	& 33751.370(1)	& 1.606	& 22.4\arcsec	& CDMS \\

HCS$^+$			& 1 - 0			& 42674.195(1)	& 0.000 & 17.7\arcsec	& CDMS \\
H$_2$CS			& 1$_{01}$ - 0$_{00}$	& 34351.43(2)	& 0.000 & 22.0\arcsec	& CDMS \\

\lneutralBrett		& J=$\frac{3}{2}$ - $\frac{1}{2}$ $\Omega$=$\frac{1}{2}$ F= 2 - 1 f	& 32627.297(2)	& 0.00086  & 23.2\arcsec & CDMS \\
\lneutralBrett		& J=$\frac{3}{2}$ - $\frac{1}{2}$ $\Omega$=$\frac{1}{2}$ F= 2 - 1 e	& 32660.645(2)	& 0.00086  & 23.1\arcsec & CDMS  \\	
\lBrett			& 1 - 0			& 22489.864(2)	& 0.00	& 33.6\arcsec	& CDMS \\
\lBrett			& 2 - 1			& 44979.544(3)	& 1.079	& 16.8\arcsec	& CDMS \\

\hline
\end{tabular}
\end{table*}
\normalsize

\subsubsection{OH and \cycloprop} 
 \label{subsubsec:profilesOHcycloprop}
\begin{figure*}
\centering
\includegraphics[width=15cm]{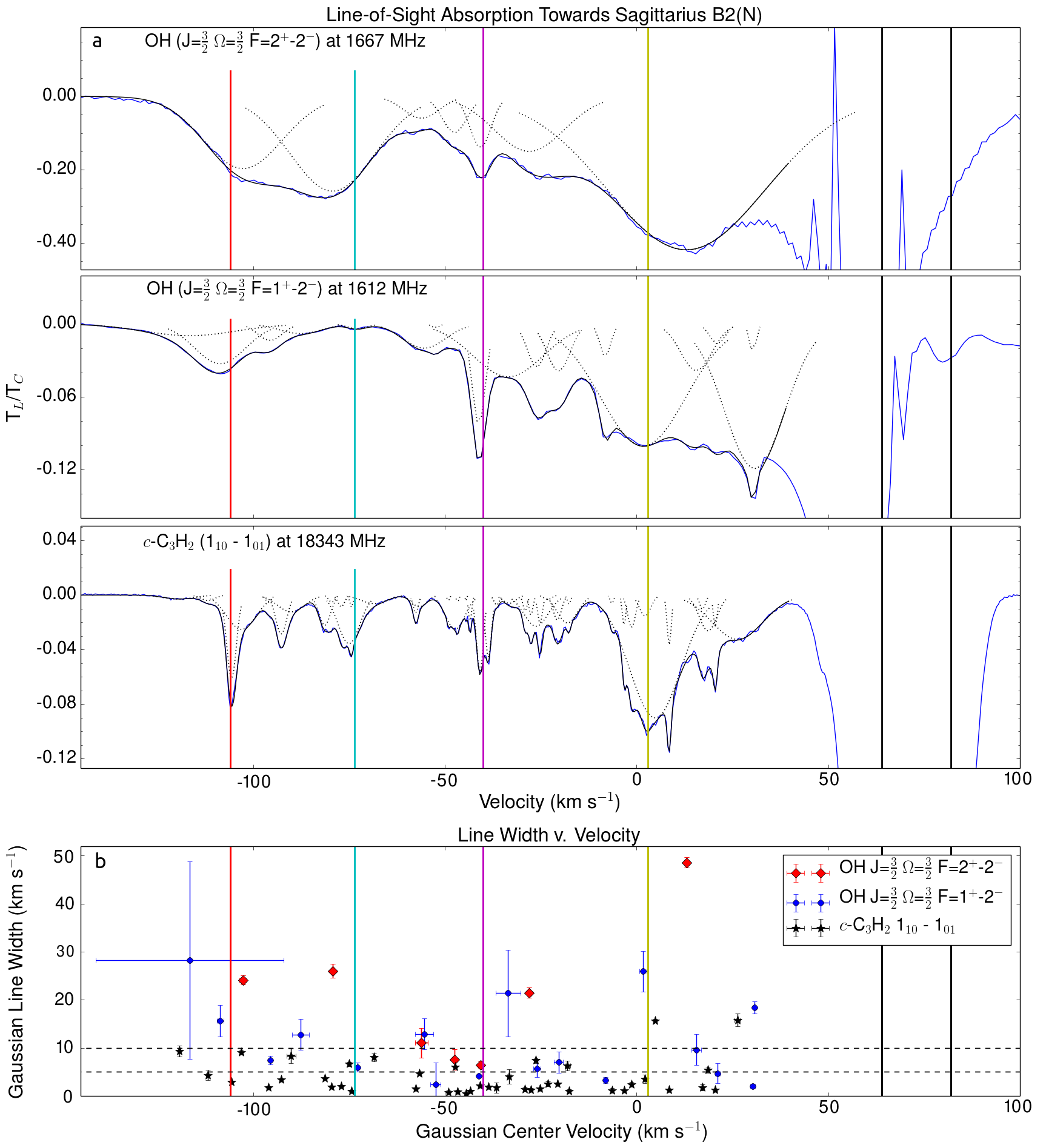}
\caption{ {\bf a.} Line-of-sight profiles of OH and \cycloprop ~are plotted in blue with individual Gaussian fit 
components shown with black dashed lines and the complete Gaussian fit overlaid in solid black.  
{\bf b.} Gaussian fit parameters of center velocity and line width for individual Gaussian components 
fit to the profiles of OH and \cycloprop.  Horizontal dashed lines indicate line widths of 5 \kms ~and 10 \kmss. 
In all panels, black vertical lines indicate the velocities of line absorption by Sgr B2 at +64 and +82 \kmss, 
and colored lines are located at -106, -73.5, -40, and +3~\kmss.}
\label{OH_cC3H2_profs}
\end{figure*}

Figure \ref{OH_cC3H2_profs} shows the absorption profiles and Gaussian component fits to selected lines of 
OH and \cycloprop.
The profile of OH (J=$\frac{3}{2}$ $\Omega$=$\frac{3}{2}$ F=2$^+$-2$^-$) at 1667 MHz
includes very broad line components, many of which have \delv ~\gsim ~20~\kmss,
and the absorption fully covers the velocity range of -120 to +40 \kms ~with no absorption-free channels. 
While the OH (J=$\frac{3}{2}$ $\Omega$=$\frac{3}{2}$ F=1$^+$-1$^-$) transition at 1665 MHz is 
extremely similar to the profile of the 1667 MHz transition (Appendix A), the 
(J=$\frac{3}{2}$ $\Omega$=$\frac{3}{2}$ F=1$^+$-2$^-$) line profile at 1612 MHz appears somewhat different.  The 
1612 MHz transition has deeper absorption by the -40 \kms ~cloud and weaker absorption in the velocity range of -90 to -60 \kmss,
as discussed in \S \ref{subsec:OHabund}.  While the line absorption at 1612 MHz also fully covers the velocity range 
of -120 to +40 \kms ~and contains broad (\delv~\gsim~10~\kmss) Gaussian components over this range, 
lower line width features, of 3~\lsim~\delv~\lsim~10~\kmss, are superimposed on the broad line absorption.
In contrast to the broad line widths of the 1667 MHz OH transition in particular, absorption by \cycloprop ~consists of comparatively narrow
features, leaving segments of the spectrum without detectable absorption.  In this profile, many 
narrow components (with \delv~\lsim~3\kmss) are superimposed on moderately broad components with 3~\lsim~\delv~\lsim~10~\kmss.


In the line absorption by \cycloprop, at least ten line-of-sight absorption 
components are clearly distinguishable, peaking at velocities from -106 to +25 \kmss.
Throughout this section, we treat these features independently and refer to these clouds by their approximate center velocities.  
The center velocities and velocity ranges considered are provided in Table \ref{tab:vranges}.
Due to the nearly constant abundance of \cycloprop ~in diffuse and translucent gas \citep{LisztcC3H2},
we use this species as a diagnostic for the total column of H$_2$ and compare the profiles of other species to \cycloprop.  
We further discuss the validity of this in \S \ref{subsec:NH2_cC3H2}.

\begin{table}[!htb]
\centering
\small
\caption{Velocity ranges considered for each cloud}
\label{tab:vranges}
\begin{tabular}{lrr}
\hline
Nominal Velocity & \multicolumn{1}{c}{$v_0$}	& \multicolumn{1}{c}{$v_1$}	 \\
(\kmss) & (\kmss) 	& (\kmss) 	 \\
\hline
-120		& -126 	& -116	\\
-106		& -110	& -97 	\\
-92		& -97	& -86 	\\
-80		& -86 	& -78  	\\
-73		& -78 	& -62 	\\
-58		& -62	& -53	\\
-47		& -53	& -42.5	\\
-40		& -42.5 & -35 	\\
-23		& -35	& -12 	\\
0 		& -12	& 15 	\\
20	 	& 15	& 24	\\
\hline
\hline \end{tabular}
\label{tab:boundingVels}
\end{table}
\normalsize

\subsubsection{\formald, CS, and SO} 
 \label{subsubsec:profiles_formaldCS_SO}
\begin{figure*}
\centering
\includegraphics[width=15cm]{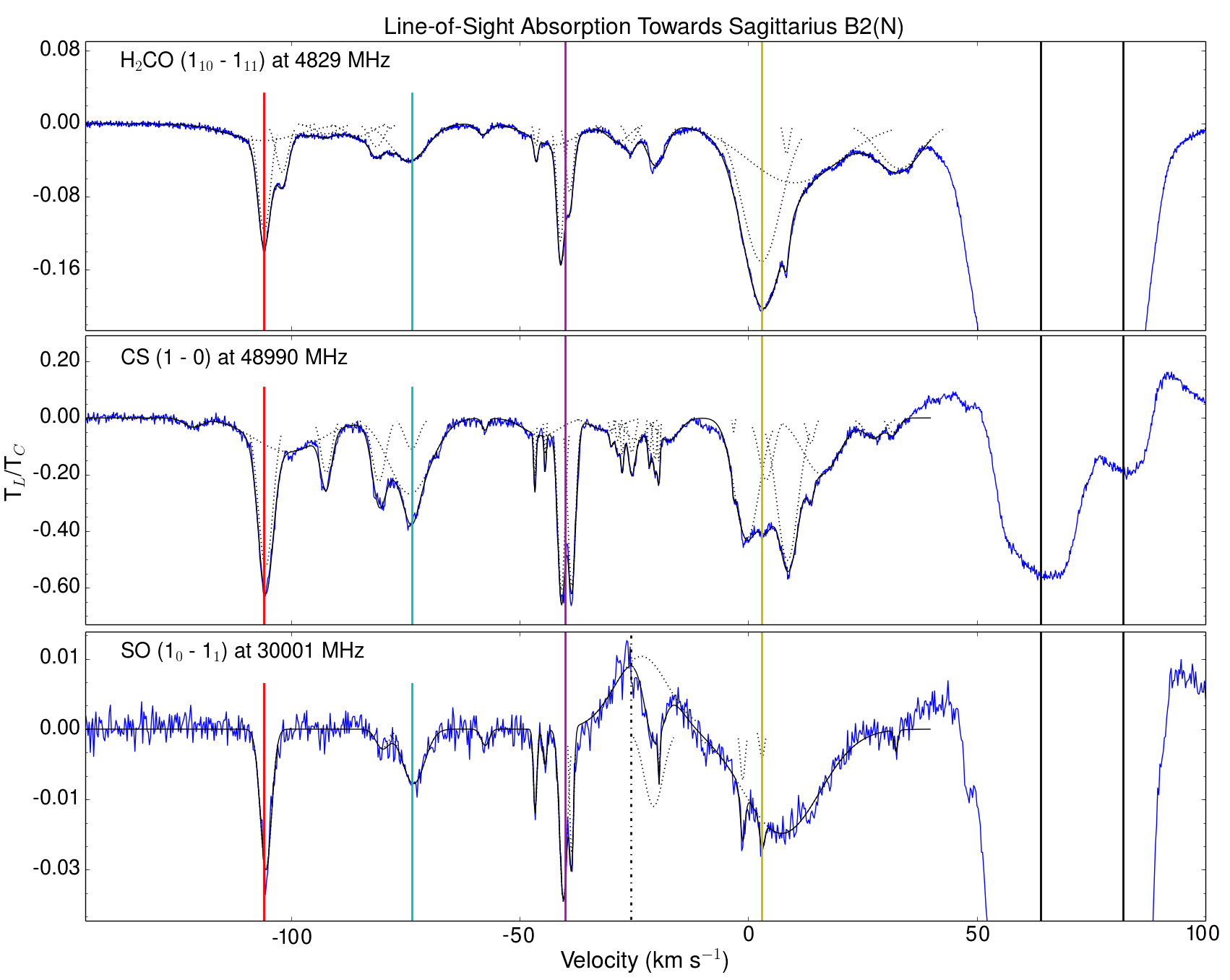}
\caption{The absorption profiles of \formald, CS, and SO are overlaid by individual best fit Gaussian components (black dotted line) 
and the sum of best-fit Gaussians (black solid line).    
Black vertical lines indicate the velocities of line absorption by Sgr B2 at +64 and +82 \kmss, and colored lines are located at -106,
-73.5, -40, and +3~\kmss. In the profile of SO, the black dashed-dotted line indicates the 8$_{17}^{\hspace{1.3mm}-}$ - 8$_{18}^{\hspace{1mm}+}$ transition of 
\meOH ~at +64 \kmss, which appears in emission.}
\label{H2CO_CS_SO_profs}
\end{figure*}

Figure \ref{H2CO_CS_SO_profs} shows high signal-to-noise transitions of \formald, CS, and SO. \formald ~and CS 
are detected in the same ten velocity components as \cycloprop, while SO is detected in all components with the exception 
of the -92 \kms ~cloud.  Line absorption by \cycloprop ~and \formald ~are fairly similar 
but have notable differences. Whereas absorption by \cycloprop ~is
of a similar peak absorption strength in the -92, -73, -40, and -23~\kms ~clouds, the 
\formald ~absorption varies more substantially in these components.  In the \formald ~profile, 
the -73~\kms ~and the -23~\kms ~components have peak absorption strengths, whereas the -40~\kms 
~component is a factor of $\sim$3 stronger and the -92~\kms ~cloud is significantly weaker.  
Additionally, whereas peak absorption by \cycloprop ~is $\sim$40~percent stronger at -106 \kms ~than in the -40 \kms ~cloud,
\formald ~has deeper absorption in the -40 \kms ~cloud than in the -106 \kms ~cloud.

The absorption profile of CS (1-0) appears qualitatively distinct from the line profiles of \cycloprop ~and \formald.  
Whereas  \cycloprop ~and \formald ~exhibit comparatively smooth absorption profiles, CS line absorption 
is characterized by sharper, more jagged features.  This is particularly true of the 
clusters of clouds around -47, -40, and -23 \kmss, where CS absorbs in narrow, sharp peaks without as much 
broad-component absorption as is present in the line profiles of \cycloprop ~and \formald.  This is best illustrated by the -47 \kms
~gas, in which the absorption depths of the narrow features are significantly greater than the strength of the broad absorption component in the CS 
profile, whereas the opposite is true for \cycloprop ~and \formald.
Similarly, absorption by SO (1$_0$-1$_1$) has narrow lines with little broad-component absorption in the -47 and -40 \kms ~clouds.

\subsubsection{CS, CCS, \HCS, and \HtwoCS} 
 \label{subsubsec:profiles_CSbeearing}
\begin{figure*}
\centering
\includegraphics[width=15cm]{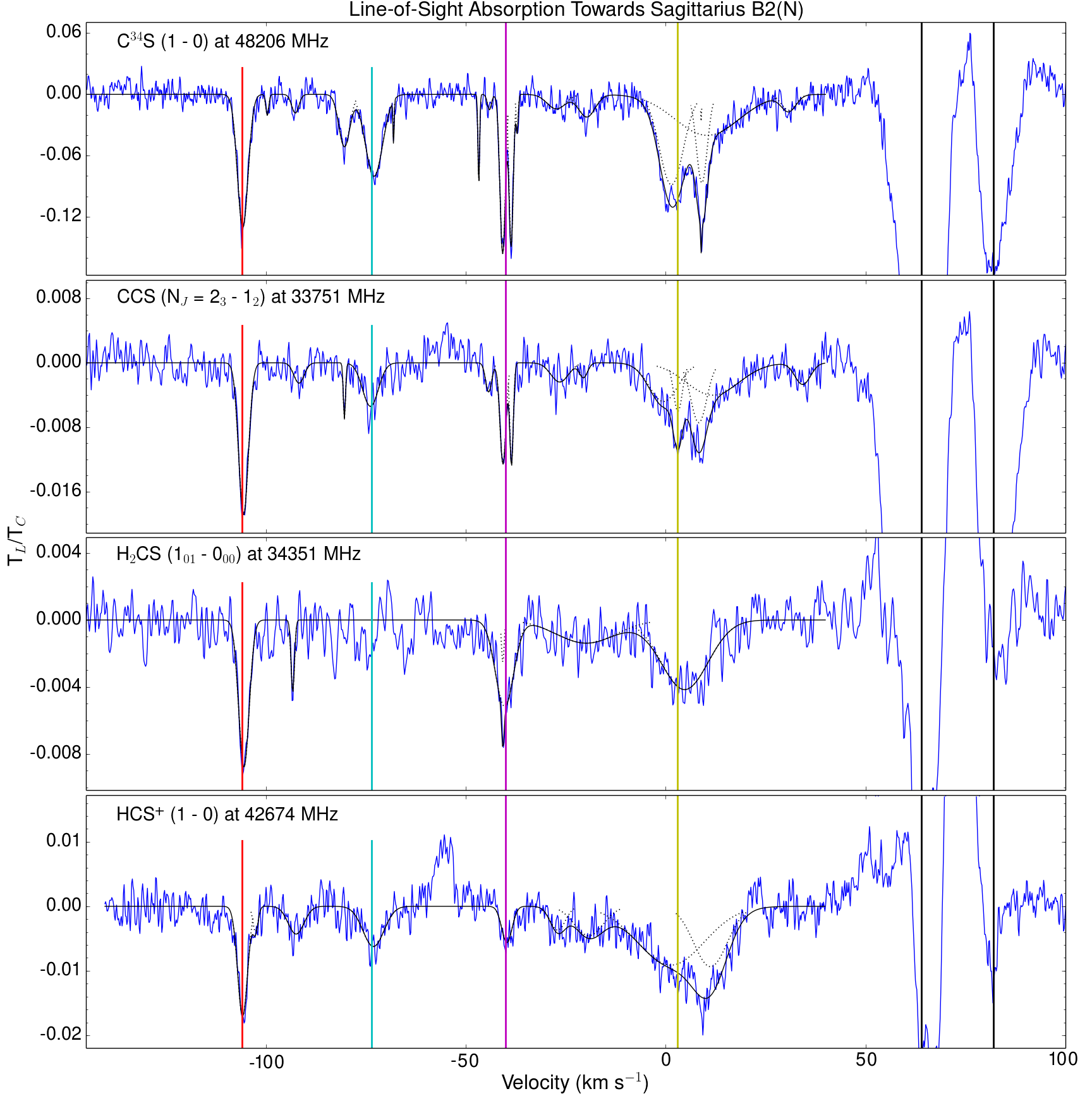}
\caption{The absorption profiles of CS-bearing species are overlaid by the best fit Gaussian components (black dotted line) 
and the sum of best-fit Gaussians (black solid line).  
Black vertical lines indicate the velocities of line absorption by Sgr B2 at +64 and +82 \kmss, and colored lines are located at -106,
-73.5, -40, and +3~\kmss. }
\label{CSbearing}
\end{figure*}

Figure \ref{CSbearing} includes selected line profiles of CS-bearing molecules. Generally, the profiles of C$^{34}$S and CCS 
are similar to one another, with strong line absorption in the \lineClouds clouds and narrow peaks 
characterizing the -47 and -40 \kms ~gas. The profiles of \HtwoCS ~and \HCS ~appear to be less characterized 
by deep, narrow absorption features, although this could be because these two profiles do not have sufficient signal-to-noise to 
permit a full characterization of narrow features. The patterns of absorption are somewhat different for these two species than for CS, C$^{34}$S, and CCS.
For \HtwoCS, little to no absorption is detected at -73 \kmss. In the profile of \HCS, only weak absorption by \HCS~is observed in the -40 \kms ~component 
whereas CS, \CSiso, CCS, and \HtwoCS ~produce strong absorption in this cloud.

\subsubsection{SiO} 
 \label{subsubsec:profiles_SiO}
\begin{figure*}
\centering
\includegraphics[width=15cm]{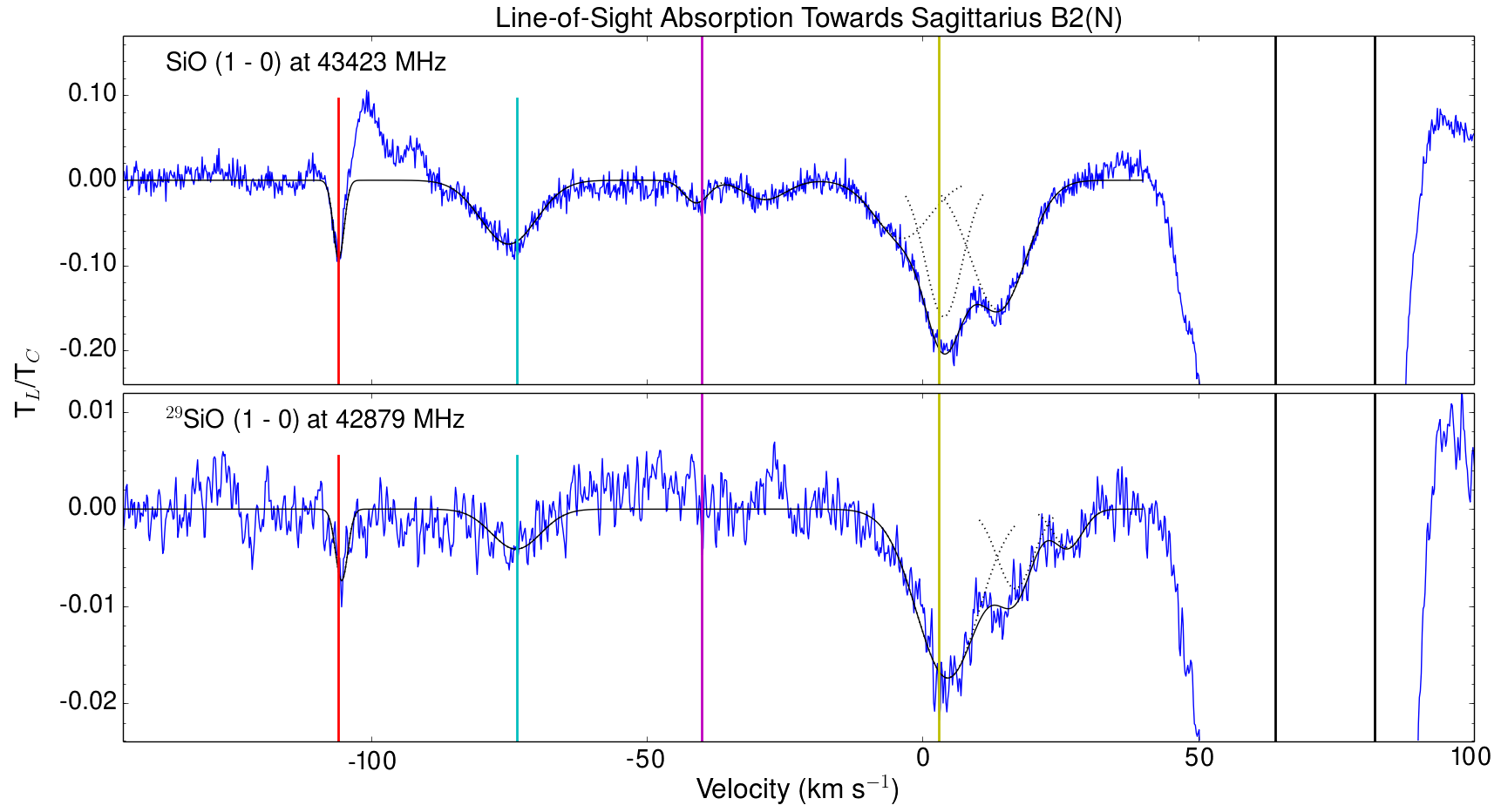}
\caption{The absorption profiles of  $^{28}$SiO and  $^{29}$SiO are overlaid by the best fit Gaussian components (black dotted line) 
and the sum of best-fit Gaussians (black solid line).  
Black vertical lines indicate the velocities of line absorption by Sgr B2 at +64 and +82 \kmss, and colored lines are located at -106,
-73.5, -40, and +3~\kmss.}
\label{SiOprofs}
\end{figure*}

Figure \ref{SiOprofs} shows the line profiles of $^{28}$SiO and $^{29}$SiO. Compared to \cycloprop, \formald, and CS-bearing species especially, 
the profiles of SiO isotopologues are much smoother and are predominantly characterized by moderately broad components (with 3~\lsim~\delv~\lsim~10~\kmss). 
Strong absorption is present particularly at -73 and 0 \kmss.  Furthermore, the clouds at -73 and -80 \kms ~do not show any distinction from one another in 
the SiO profile;  whereas the two clouds have separate peaks in other transitions, they form a single feature in the SiO profile.

\subsubsection{\lneutralBrett ~and \lBrett} 
 \label{subsubsec:profiles_lC3H}
\begin{figure*}
\centering
\includegraphics[width=15cm]{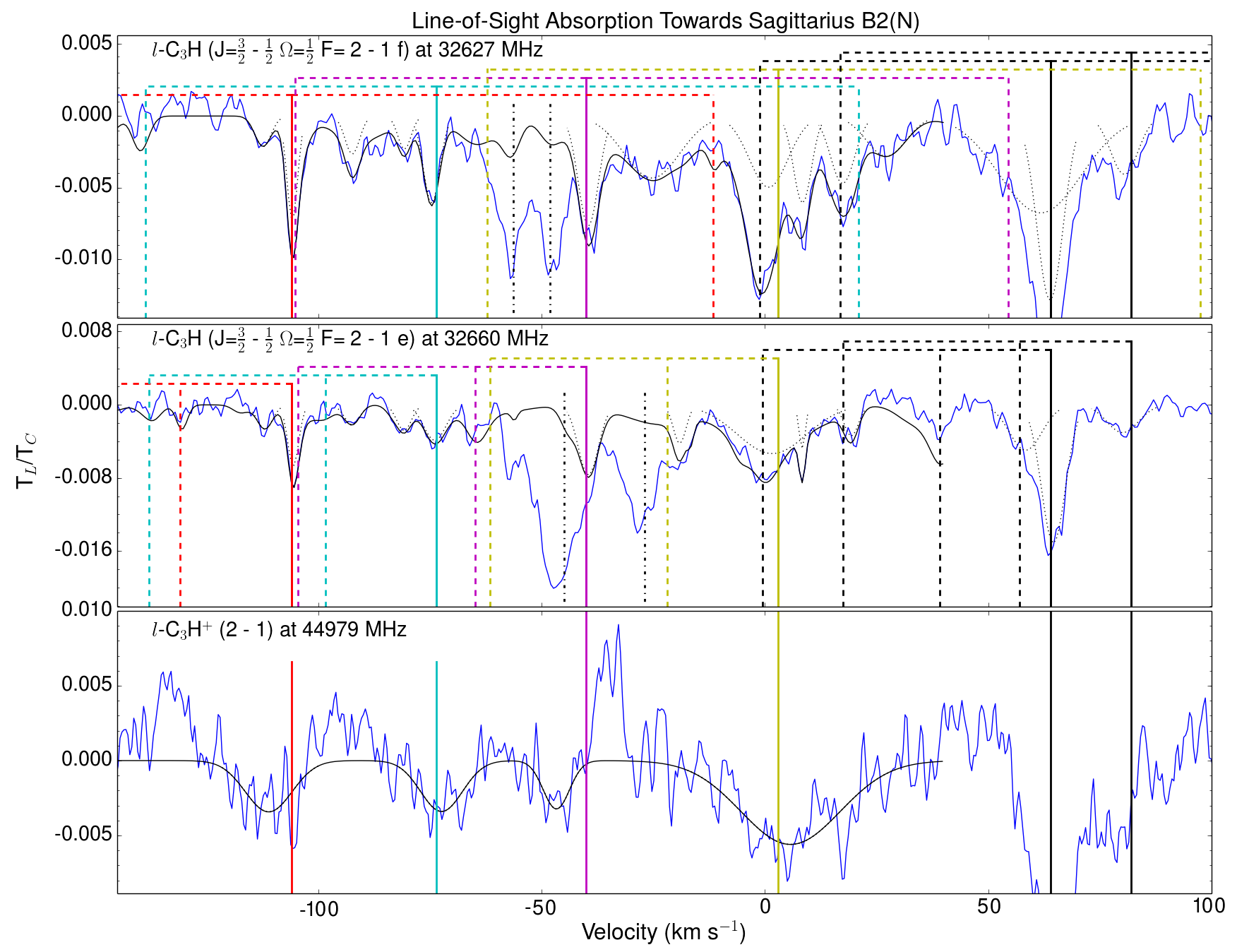}
\caption{Absorption profiles of \lneutralBrett ~and \lBrett ~are overlaid by the best fit Gaussian components (black dotted line) 
and the sum of best-fit Gaussians (black solid line).  
Black vertical lines indicate the velocities of line absorption by Sgr B2 at +64 and +82 \kmss, and colored lines are located at -106,
-73.5, -40, and +3~\kmss.  Hyperfine structure for each of these velocity components is indicated by dotted vertical lines of the same color.
The data are overlaid by Gaussian components fit to the main hyperfine component shown in black dotted lines, and by the total 
fit to the profile, which assumes that hyperfine or A/E components are present with the same line shape as the primary component, but 
with the height scaled by the ratio of the line strengths.
In the profile of \lneutralBrett ~at 32627 MHz, the black dashed-dotted line marks unidentified transitions that are not consistent with the 
typical profile of diffuse cloud absorption in this line of sight.  In the line profile of \lneutralBrett ~at 32660 MHz, the 
black dashed-dotted line marks a transition of \cisC ~at velocities of +64 and +82 \kms ~associated with Sgr B2. }
\label{lC3Hprofs}
\end{figure*}

The line profiles of \lneutralBrett ~and \lBrett ~are shown in Figure \ref{lC3Hprofs}.  
As the transitions of \lneutralBrett ~have hyperfine structure, we have only fit the primary (F=2-1) line, with 
these components shown by dashed Gaussian profiles, and we add the hyperfine structure to the total fit (solid black line) as described above.  
The presence of the hyperfine structure, combined with the somewhat weak signal and line blending, makes it difficult to 
characterize the line profiles of \lneutralBrett.  
From the (J=$\frac{3}{2}$\,-\,$\frac{1}{2}$ $\Omega$=$\frac{1}{2}$ F\,=\,2\,-\,1 $l$=f) line profile at 32\,627 MHz, we can confidently
characterize the absorption at -106, -92, -80, -73, -40, -23, and 0 \kmss.  In this line profile, 
we also notice strong absorption at -57 and -48 \kmss, possibly arising from 
the -58 and -47 \kms ~clouds. In the line-of-sight profiles by other molecules in the PRIMOS data, however, absorption by the -58 and -47 \kms ~clouds 
is weaker than absorption by the -40 and -73 \kms ~clouds.  Thus, if the absorption at -57 and -48 \kms ~is from \lneutralBrett,  then 
the -58 and -47 \kms ~clouds would have exceptionally high abundances of \lneutralBrett ~compared to other clouds.
If the features are from \lneutralBrett, the (F\,=\,1\,-\,0) hyperfine transitions should occur at -122 and -113 \kmss, respectively,
in the rest frame of the (F\,=\,2\,-\,1) transition.  Absorption features are present at both of these velocities, although the observed features are 
somewhat weaker than the hyperfine lines should be given the strength of the (F\,=\,2\,-\,1) features. 
In the ($l$\,=\,e) transition at 32660 MHz, the velocity range of -60 to -20 \kms ~is obscured by blending with line absorption 
from {\it cis-}glycolaldehyde (\cisC) ~in Sgr B2 at +64 and +82 \kmss.  Nonetheless, an absorption feature at -58 \kms ~is possibly detectable and 
somewhat resolved on the low-velocity edge of the +64 \kms ~component of \cisC.   This profile is also not 
inconsistent with there being line absorption by \lneutralBrett ~at -47 \kmss, although it is not possible to claim this feature is detected.
However, if absorption by the (F\,=\,2\,-\,1 $l$\,=\,e) transition is present at -58 and -47 \kmss, the (F\,=\,1\,-\,0) hyperfine components 
would occur at -80 and -73  \kmss, and would account for all absorption at these velocities.  Because the -73 and -80 \kms ~components were detected
in the ($l$\,=\,f) component, we are inclined to believe that the absorption at -80 and -73 \kms ~is from the -80 and -73 \kms ~clouds, and are not hyperfine 
features from the -58 and -47 \kms ~clouds.  Therefore, we treat the features at -48 and -57 \kms ~in the ($l$\,=\,f) profile at 32627 MHz as unidentified
lines but warn of the possibility that the absorption is from the line-of-sight clouds.

In the line profile of \lBrett ~(2\,-\,1), line absorption is clearly detected within the 0 \kms ~cloud, and absorption is also present in the -106, -73, 
and -47 \kms ~clouds.  An unidentified emission line at 44994 MHz in the rest frame of Sgr B2 prevents characterization of the -40 \kms
~cloud.  In the (1-0) transition of \lBrett ~(Appendix A), absorption is also clearly detected at \zero \kmss, and the profile provides 
evidence supporting the detection of \lBrett ~in the -73 \kms ~cloud.  The detection of \lBrett ~in the -106 \kms ~cloud remains somewhat suspect, as it 
either contains substantial absorption from -120 to -110 \kms ~or is confused with a different line or with baseline effects.  
We note that the baseline is stable on either side of the velocity range of interest, so do not favor the latter explanation.  
However, as the -106 \kms ~component is too weak to be detected in the (1-0) transition,
we treat \lBrett ~in the -106 \kms ~cloud tentatively, and recommend that it should be confirmed with an additional transition.

\subsection{Cloud Kinematics}  
\label{subsec:kinematics}
\begin{figure*}[!htb]
\centering
\includegraphics[width=15cm]{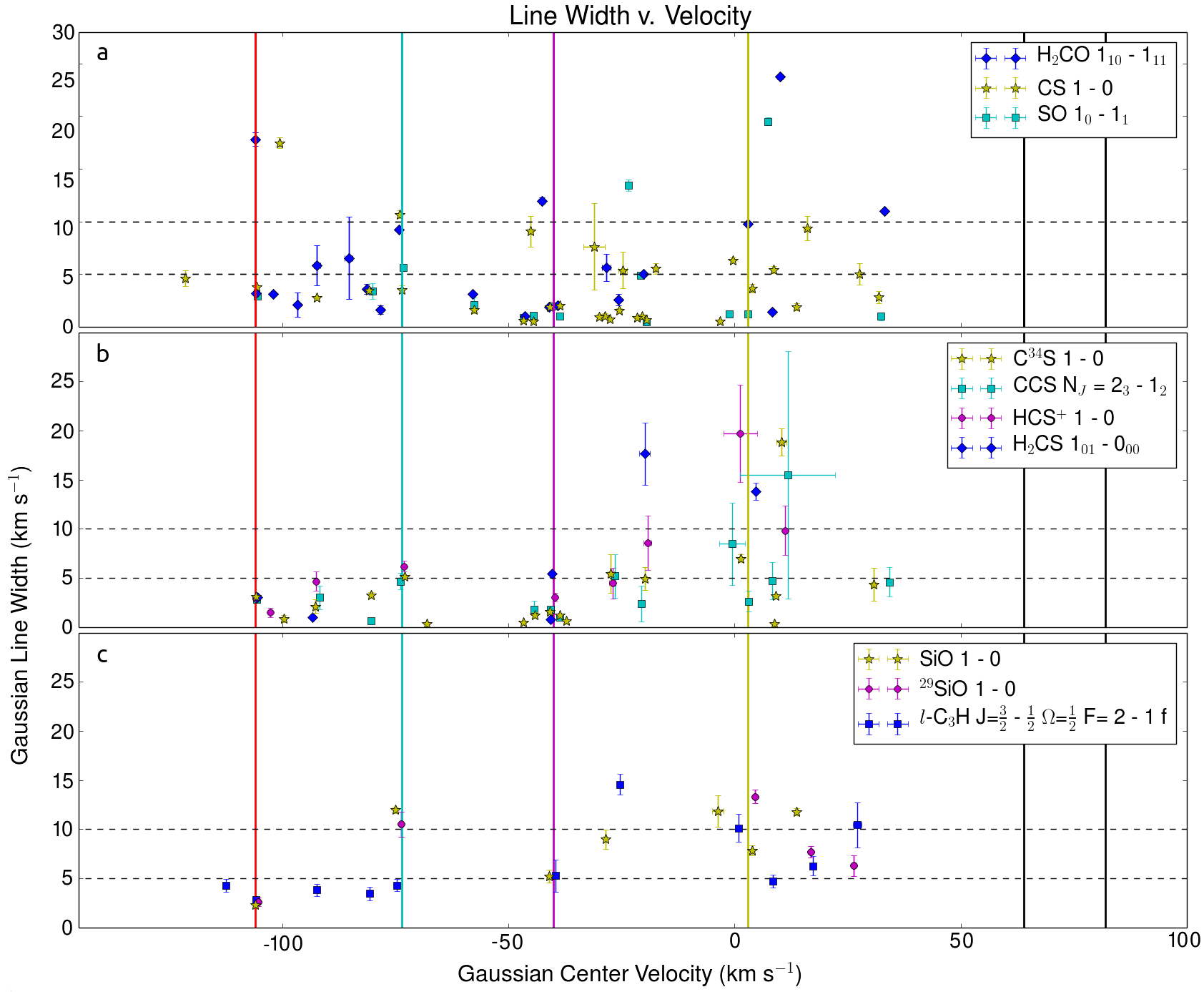}
\caption{Gaussian fit parameters of center velocity and line width for individual Gaussian components 
fit to selected transitions of {\bf a} \formald, CS, and SO, {\bf b} CS-bearing molecules, and {\bf c} isotopologues of SiO and \lneutralBrett.  
In all panels, black vertical lines indicate the velocities of line absorption by Sgr B2 at +64 and +82 \kmss, 
and colored lines are located at -106, -73.5, -40, and +3~\kmss.
Horizontal dashed lines indicate line widths of 10 \kms ~and 5\kmss. }
\label{centerLW_combined}
\end{figure*}

Figure \ref{centerLW_combined} shows the Gaussian fit parameters of center velocity and line width for all
components fit to selected line profiles.  
In this figure, it is apparent that the -106 \kms ~component has particularly consistent and 
well defined parameters of line center and width, at $v \approx$ -105.7 \kms ~and $\Delta v \approx$ 3.1 \kmss.
The standard deviations on these parameters, of 0.2 and 0.3, respectively, are consistent 
with the errors to individual measurements. Although the profile of \formald ~(1$_{10}$-1$_{11}$) includes a weak
broad component at or near -106 \kmss, this line and all other 
transitions shown in Figure \ref{centerLW_combined} have absorption at ($v$, $\Delta v$) $\approx$ (-105.7 \kmss, 3 \kmss).
For some high signal-to-noise transitions, a narrow wing is present to the positive velocity side of the -106 \kms ~cloud, 
at approximately -102 \kmss, and a moderately broad wing is present towards the negative velocity side, near -115 \kmss.

The -92 \kms ~cloud is centered at -92.6 \kms ~and has mean width of $\Delta v$ = 2.7 \kmss.  
The line widths vary significantly more for the -92 \kms ~cloud than for the -106 \kms ~cloud, 
with measured values ranging from $\Delta v =$ 1 to 5 \kmss.  

In the -80 \kms ~cloud, most line fits are tightly clustered around 
$v \approx$ 80.4 \kms ~and $\Delta v \approx$ 3.5 \kmss.  Absorption by the  -73 \kms ~cloud typically has a larger line width, 
with a median value of 4.9 \kmss.  Whereas lines of SiO, and \formald 
~have higher line widths, of 9-12 \kmss, most other species have widths of $\sim$5 \kmss, although significant scatter exists. 

The -58 \kms ~cloud is centered at -57.6 \kmss, and has a line width of $\sim$2.1 \kms ~with statistically significant scatter.

The -47 \kms ~cloud consists of two narrow velocity components, at -46.6 and -44 \kmss, in addition to a broad component 
that is prominent in transitions of \cycloprop ~and \formald.  The narrow components have very consistent line center velocities, and 
typical widths of 0.7 and 1.0 \kms ~at -46.6 and -44 \kmss, respectively.  Similarly, for small species with high signal-to-noise, the 
-40 \kms ~component consists of two narrow lines, at -40.8 and -38.7 \kmss, with widths of 1.9 and 1.5 \kmss, respectively.  For lower 
\StoN ~transitions, the two components are fit with a single Gaussian with a higher line width.  

As observed in the highest signal-to-noise transitions, the -23 \kms ~cloud consists of many narrow features superimposed 
on two moderately broad features which dominate the integrated flux.  
These two moderately broad features are at -27.1 and -20.1 \kmss, with widths of 4.8 and 4.6 \kmss, respectively, 
although significant scatter exists.

Finally, the \zero and +20 \kms ~clouds contain multiple superimposed components, including broad profiles (\delv~\textgreater~10~\kmss) with moderately broad
(3~\textless~\delv~\textless~10~\kmss) and narrow (\delv~\textless~3~\kmss)
features superimposed.  Fits to this material typically include a feature at -1.8 \kmss, a strong component at +3.5 \kmss, 
a component at +9 \kmss, a broad component centered between 10~\textless~v~\textless~15 \kmss, and a broad feature at +30 \kmss.  
The exact profiles vary significantly in different lines however, producing substantial scatter in Figure \ref{centerLW_combined}.

\subsection{Molecular Column Density Measurements} 
\label{subsec:colDenssAbunds}

We determined column densities of every molecule in each distiguishable cloud component.
To do so, we obtained the integrated line optical depths by 
\begin{multline} \label{eq:intOD}
 \int \tau_l dv = \int_{v_0}^{v_1} -ln \left( 1 + \frac{T_L}{T_C - [f(T_{Ex}) - f(T_{\text{CMB}})]}\right);\hspace{1cm} \\
 f(T) = \frac{h\nu/k}{e^{h\nu/kT}-1} \leq T \hspace{1cm}
\end{multline}  
\citep{Nyman84,GN96} with \TCMB~= 2.73~K over the bounding velocities $v_0$ and $v_1$~provided in Table \ref{tab:boundingVels} for clouds at
-120, -106, -92, -80, -73, -58, -46, -40, -23, $\sim$0, and +20~\kmss.  We convert from integrated line optical depths to 
column densities by: 
\begin{equation} \label{eq:colDens}
N = 8.0 \times 10^{12}\, {\frac{Q}{S_{ij}\mu^2} }\hspace{1.5mm} { \frac{\int \tau dv}{e^{-E_L/kT_{Ex}} - e^{-E_U/kT_{Ex}}} }\hspace{1.5mm} \text{cm}^{-2}\hspace{1cm}
\end{equation}  
by the conventions used in \citet{LL93} and \citet{GN96}, where Q is the partition function, $E_L$ and $E_U$ are the upper and 
lower state energies, $S_{ij}$ is the intrinsic line strength, and $\mu$ is the transition dipole moment in Debye.
The calculation presumes that the absorbing gas extends homogenously over the continuum.  This assumption 
is standard in studies of diffuse and translucent gas in line-of-sight absorption \citep[e.g.][]{GN96,LL_CS,Qin2010}.

While the value of the partition function is typically estimated from equations like those provided in \citet{McDowell, McDowell2}, 
these equations diverge from the true values at low excitation temperatures of \Tex \textless 10~K. 
Due to the extremely low excitation temperatures in the line-of-sight clouds,
we determined the partition functions directly by counting states with $Q = \Sigma g e^{-E/kT_{ex}}$.
For this computation and for input line parameters required in Equation \ref{eq:colDens}, line data 
from the Cologne Database of Molecular Spectroscopy \citep[CDMS;][]{CDMS} and the 
NASA Jet Propulsion Laboratory catalogue \citep[JPL;][]{JPL} 
were accessed through the ALMA Spectral Line Catalogue\,-\,Splatalogue\footnote{The ALMA Spectral Line Catalogue is available at www.splatalogue.net; \citep{Remijan07}}.
In the data output from the CDMS and JPL catalogs, the value \SijU ~incorporates the degeneracy of the upper state $g_U$.
The value of the degeneracy varies depending on how states are defined, i.e. whether hyperfine states and torsionally excited states
are treated individually or collapsed. As a result, care must be taken to ensure consistency between 
the adopted or calculated value of the partition function and the line parameters used for the column density calculation.  
In our work, the partition functions were computed directly from the line data at $T$ = 3~K using all states with energy 
$E$~\textless~100~K (and up to 300~K for molecules with fewer transitions).  
We further verified that our calculated values of the partition function are in agreement with the values published
by the catalogs at temperatures of 9.375, 18.75, and 37.5 K. 

We determined errors to the integrated line optical depth measurements by including error contributions from three sources added in quadrature.  
These include (1) error to the integrated line optical depth
estimated by the root mean squared residual between the line profile fit and the data times the width of the velocity range considered, 
(2) a 2.5 mK baseline uncertainty, which is a typical level of variation in the baseline offset within a single line profile, and 
(3) an estimated 5 percent uncertainty in the denominator of Equation \ref{eq:intOD}.

\begin{figure*}[!htb]
\centering
\includegraphics[width=16.5cm]{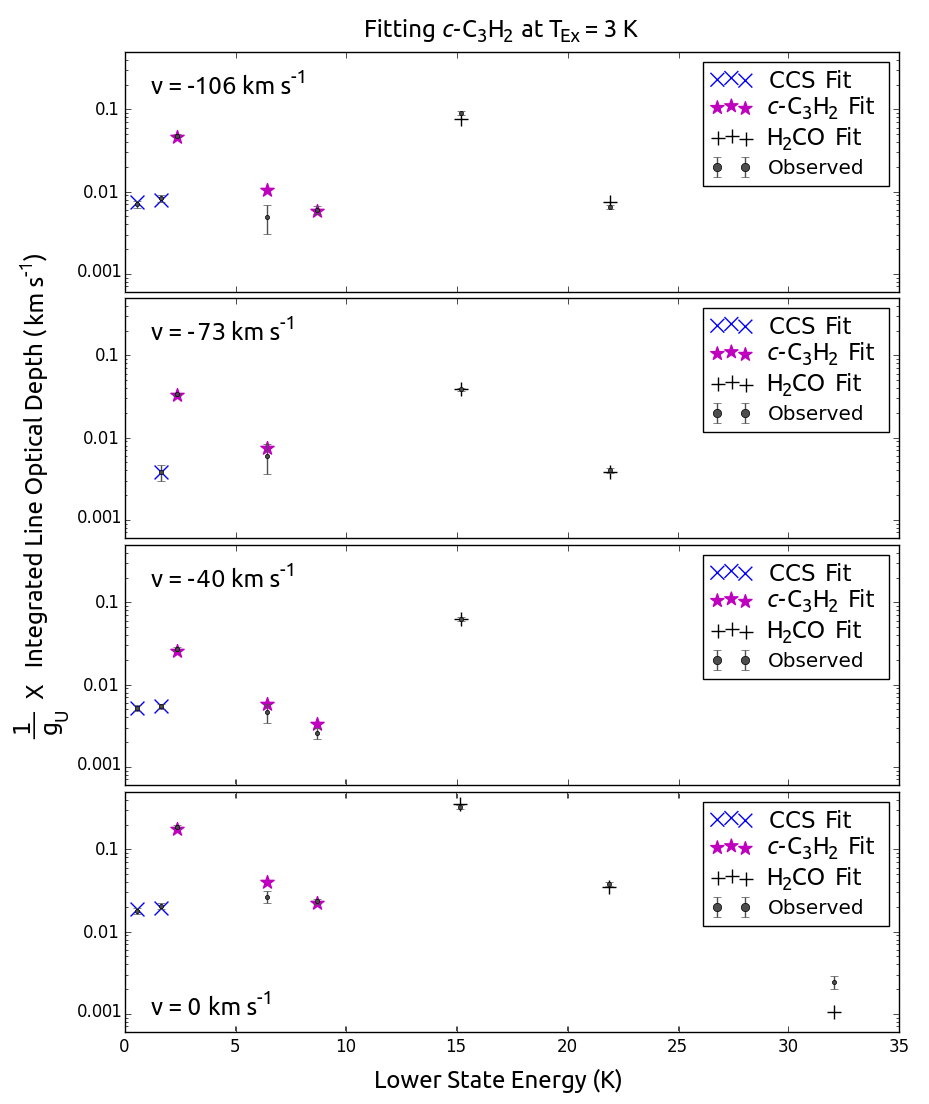}
\caption{Observed integrated line optical depths and best fit models, computed at \Tex = 3K, of CCS, \cycloprop, and \formald.  
Error bars indicate 1$\sigma$ errors. }
\label{ColDensFits}
\end{figure*}

We determine that a model with \Tex\,=\,3~K does a good job of accounting for the line absorption for nearly 
all molecules in nearly all components.  Initially, for molecules with multiple transitions in our data set, 
we allowed both \Tex and \Nmol ~to vary and determined that values 
of 2.5\,-\,3.7~K produced the best fits. After determining that 3~K is appropriate, we set \Tex = 3~K 
before obtaining the best-fit values for the molecular column density. For illustration, Figure \ref{ColDensFits} shows the 
modeled and observed values of the integrated line optical depth obtained with \Tex = 3~K for observed 
transitions of CCS, \cycloprop, and \formald.  
For nearly all transitions, the best fit was within 2$\epsilon$ of the 
observed value.  Further, for most molecules with multiple transitions (namely OH, CCS, \lneutralBrett, \lBrett, \cycloprop, and \formald), 
the values of \Nmol ~estimated from different transitions vary by \lsim\,10 percent, and somewhat larger variation is observed for a few 
velocity components of \cycloprop~ and \formald.

For molecules with only a single transition observed in \LOS ~absorption (i.e. SiO, SO, and CS-bearing species),
it was then straightforward to determine the molecular column density by Equation \ref{eq:colDens} with the assumption of \Tex = 3~K.  
The column density errors reported include the error from the integrated optical depth measurement as described above, and do not include 
uncertainty in the excitation temperature.

For molecules with multiple transitions, we determined the reported column density as the best fit to all measured 
transitions as determined by an error-weighted least-squares method.  The errors for individual transitions 
input to the column density fitting procedure include the three contributions specified above.  For column density estimates 
determined with multiple transitions, it was possible to estimate the total error by two methods.  The first method includes errors to the 
individual fits and estimates that the column density fit error will decrease with a greater number of constraints, by 
\begin{equation} \label{eq:error_method}
 \epsilon_\text{N}^2 = \left({\sum}~\frac{1}{\epsilon_i^2}\right)^{-1}
\end{equation}  
for the error to the column density $\epsilon_\text{N}$ and errors to individual fits $\epsilon_i$.  
The second method is estimated by measuring the agreement between the predicted and observed integrated optical depths, weighted by 
the errors on individual transitions, by 
\begin{equation} \label{eq:error_fitmethod}
 \epsilon_\text{N}^2 = {\sum} \left( \frac{1}{\epsilon_i^2} \times \left(\frac{\text{O}_i-\text{P}_i}{\text{P}_i}\right)^2\right)  
\end{equation}  
for the error to the column density ($\epsilon_\text{N}$), errors to individual fits ($\epsilon_i$), 
observed integrated line optical depths (O$_i$), and predicted integrated line optical depths 
from the best fit model (P$_i$). If the errors estimated 
for the individual transitions are appropriate, then the two methods should return similar values. 
We generally observe good agreement between estimates made by these two methods, and we report the larger 
value estimated by the two methods.  The best fit molecular column densities and error estimates 
are provided in Appendix A, and most column density errors are of order 10 percent.

\subsection{Hydrogen Column Density Measurements}
\label{subsec:hydrogenColumns}

In order to determine abundances relative to \Htwo, we estimated the hydrogen column in each component assuming that the abundance of \cycloprop 
~is constant and equal to {\it X} = ${\frac{N_{{\it c\text{-}}C_3H_2}}{N_{H_2}}}$ = 2.5 $\times 10^{-9}$.  
This assumption is based on the work of \citet{LisztcC3H2}, which found that {\it X}$_{c\text{-C$_3$H$_2$}}$ = (2-3) $\times$ 10$^{-9}$ 
with little variation in diffuse clouds in the Galactic disk, 
and in additional studies of hydrocarbon ratios in diffuse and translucent clouds throughout the Galaxy \citep{LL2000,Sheffer2008,Gerin2010b,Gerin2011}.
In \S \ref{subsec:NH2_cC3H2} we argue that this is the best method for estimating the molecular hydrogen column in the PRIMOS data.  
The resulting hydrogen columns are provided in Table \ref{tab:NHs}.  
The hydrogen columns estimated by conversion from \cycloprop ~column densities are, for most clouds, similar to previous estimates 
made in the \los ~towards Sgr B2(M) located $\sim$45 arcsec south. 
Errors to the molecular hydrogen column come from the estimated errors to the column densities of \cycloprop, and do not include possible 
variation in the abundance of \cycloprop.

\begin{table*}[!htb]
\centering
\small
\caption{Molecular and neutral hydrogen column density estimates in units of $10^{21}$ \cmtwo}
\footnotesize{
\label{tab:NHs}
\begin{tabular}{ll|rrrrrrrrrrr}
\hline
Parameter 	& Position	& \multicolumn{11}{c}{Absorption Cloud Velocity (\kmss)} 		\\
		&		& -120	&-106	&-92	& -80	&-73	&-58	&-47	&-40	&-23	& 0 	& +20 \\
\hline
N(H$_2$) 	& (N)		&0.17(2)& 3.2(2)&1.72(8)&1.12(6)&2.3(1)	&0.52(3)&1.4(1)	&1.8(2)	&3.5(4)	& 12(1)	& 3.2(5) \\
N(H$_2$)$^{[1]}$&(N)		&\multicolumn{3}{c|}{4.8(5)}&\multicolumn{2}{c|}{2.9(1)}&0.48(3)&\multicolumn{2}{c|}{3.0(1)}&3.1(4)& \multicolumn{2}{|c}{14.6(16)}  \\
N(H$_2$)$^{[2]}$ &(M) 		& \multicolumn{3}{c|}{6}&\multicolumn{2}{c|}{1.2}& 0.9&\multicolumn{2}{c|}{9}&3 &\multicolumn{2}{|c}{16}\\
N(H$_2$)$^{[3]}$ &(M)		&\multicolumn{3}{c|}{5} & \multicolumn{2}{c|}{} & 	&\multicolumn{2}{c|}{9}&4  & \multicolumn{2}{|c}{14} \\
\hline
N(\HI)$^{[4]}$ 	&(N)		&\multicolumn{5}{c|}{1.0(3)} 	 &\multicolumn{3}{c|}{1.5(6)} 	&\multicolumn{1}{c|}{4.0(6)}& &\\
\hline \hline
$f$(\Htwo)	&(N)		&\multicolumn{5}{c|}{0.94(2)}  &\multicolumn{3}{c|}{0.84(5)} 	&\multicolumn{1}{c|}{0.65(3)}& &\\
\hline \hline 
\AV$_o\,^{[5]}$	$\approx$&(N)	& 0.08 & 1.5	& 0.8	& 0.5	& 1.1	& 0.28	& 0.76	&0.96	& 2.5 	& \textgreater4.2 & \textgreater2.4	  \\
\hline \hline 

\multicolumn{12}{l}{\footnotesize{[1] Measurements are computed from PRIMOS observations in the velocity ranges listed in \citet{GN96}. }} \\ 
\multicolumn{12}{l}{\footnotesize{[2-3] Values towards Sgr B2(M) from [2] \citet{GN96} by conversion from H$^{13}$CO$^+$ and [3] \citet{Irvine87} by}}\\
\multicolumn{12}{l}{\footnotesize{conversion from $^{13}$CO.}}\\
\multicolumn{12}{l}{\footnotesize{[4] Values towards Sgr B2(N) from \citet{Indriolo2015} from 21 cm \HI absorption. }} \\ 
\multicolumn{12}{l}{\footnotesize{[5] Central extinction of the cloud, estimated by 2\AV$_o$ = \AV ~= $(N_{HI} + 2$\NHtwo)/($2.2 \times 10^{21}$ \cmtwo), from \citep{AV_NH}.} }\\ 
\hline 
\end{tabular}}
\end{table*}
\normalsize

For the sake of comparison, we 
also include the molecular hydrogen columns measured by previous authors towards Sgr B2(M), located 45 arcsec south of (N), by conversion from H$^{13}$CO$^+$~and $^{13}$CO \citep{GN96,Irvine87}.  
Although clouds are present at approximately the same velocities in the two sightlines, the relative absorption strengths in the different kinematic components vary.  
As such, our estimates of the molecular hydrogen column density are not expected to match perfectly.  However, we do see approximate agreement for most of the clouds, 
providing one indication that our conversion from \Ncycloprop~to \NHtwo~is appropriate.  

Additionally, Table \ref{tab:NHs} includes the neutral hydrogen column densities reported by \citet{Indriolo2015} in the sightline to Sgr B2(N).  These 
were measured by \HI absorption in the 21-cm line observed with the Effelsberg 100-m telescope \citep{WinkelHI}.
While \citet{Indriolo2015} reported column densities over velocity ranges that 
include multiple distinct clouds as observed by \cycloprop, dramatic trends are present in the molecular fractions measured. The molecular 
fraction in the -23 \kms ~cloud, located in the disk, is 0.65, somewhat larger than typical values of 0.4 measured in other diffuse clouds that 
contain heavy molecules \citep{LisztcC3H2}.
In the velocity range dominated by gas in the 3~kpc arm, $\sim$85 percent of the hydrogen is in molecular form, and an even higher molecular fraction 
is measured in the velocity range dominated by Galactic Center gas, with \textgreater90 percent of the hydrogen in molecular form.  
In the final row of Table \ref{tab:NHs}, we estimate the central extinctions in individual clouds by combining the 
\Htwo ~column density measurements reported in the first row with \HI column densities estimated using the approximate molecular fractions.
The total extinction in the cloud is estimated with \AV = 1 corresponding to 
\begin{equation} \label{eq:AVestimates} 
{\text N}_{HI} + 2\,{\text N}_{H2} = 2.2 \times 10^{21}~{\text \cmtwo}
\end{equation}  
from \citep{AV_NH}, and the central extinction (\AV$_o$) is half the total extinction.  The extinction estimates should be adopted as approximate values, 
as the true extinction within the clouds depends on multiple factors as discussed in \S \ref{subsec:divisions}.

\subsection{Abundance Patterns} 
\label{subsec:abundancePatterns}

\begin{figure*}[!h]
\centering
\includegraphics[width=16.5cm]{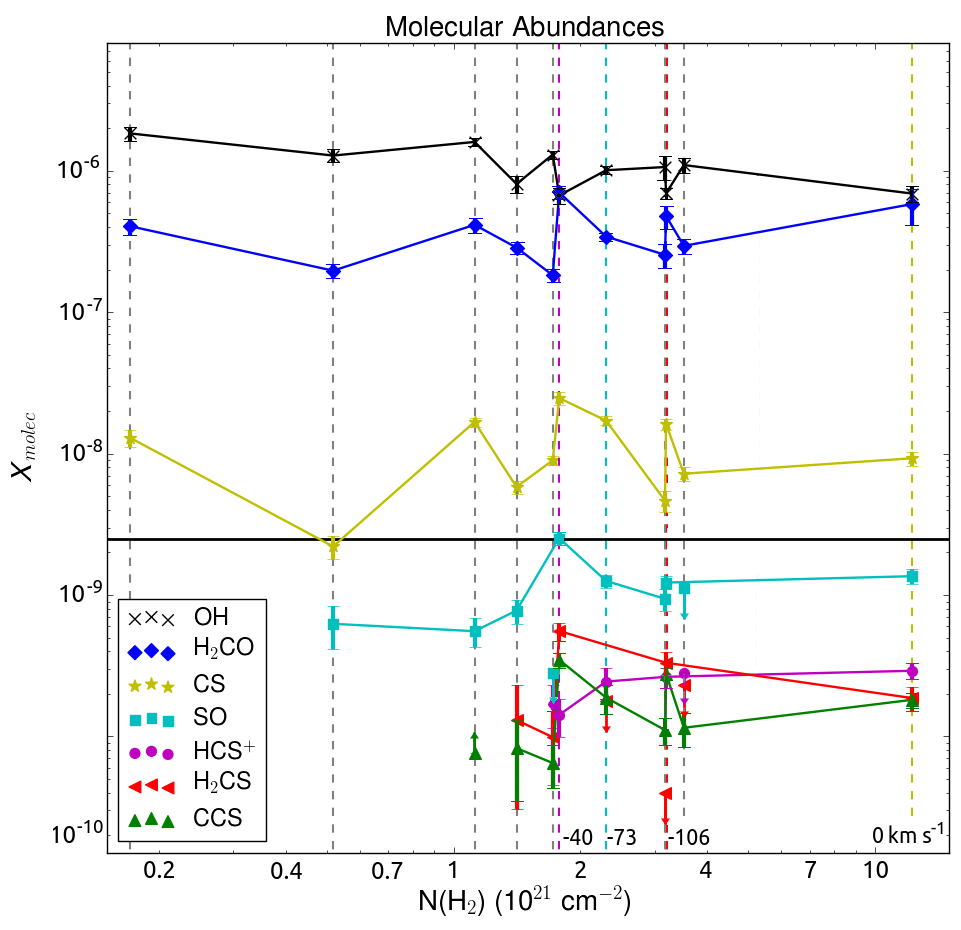}
\caption{Molecular abundances of OH, \formald, SO, and CS-bearing species.  The black solid line represents the assumed 
abundance of \cycloprop ~used for the abundance determination. In order from lowest to highest 
hydrogen column, the -120, -58, -80, -47, -92, -40, -73, +20, -106, -23, and $\sim$0 \kms ~components are marked with vertical dashed lines.  
Colored vertical lines, as labeled, mark the \lineClouds ~clouds.}
\label{Abunds_smalls}
\end{figure*}

\begin{figure*}[!h]
\centering
 \includegraphics[width=16.5cm]{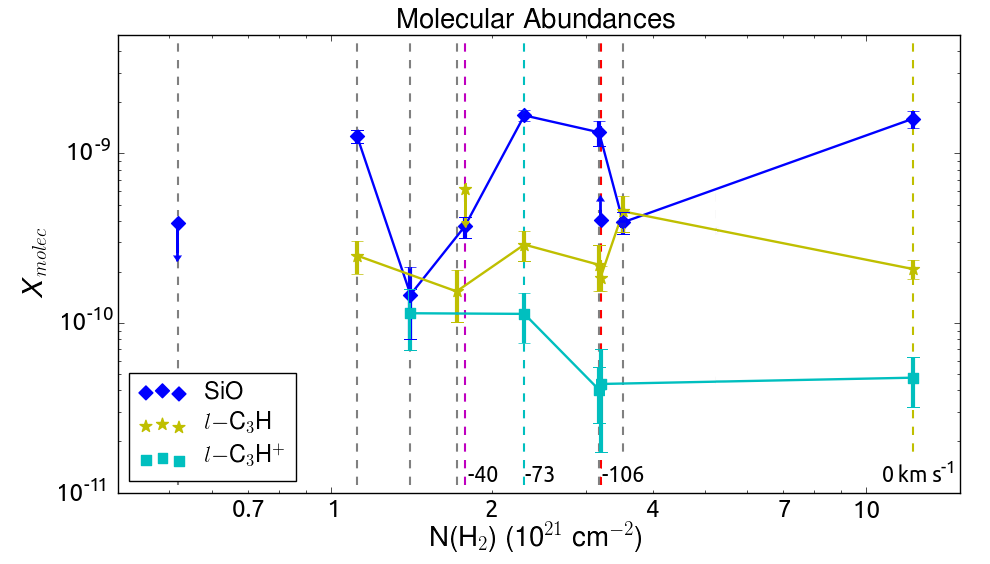}
\caption{Molecular abundances of SiO, \lneutralBrett, and \lBrett.  In order from lowest to highest 
hydrogen column, the -58, -80, -47, -92, -40, -73, +20, -106, -23, and $\sim$0 \kms ~components are marked with vertical dashed lines.
Colored vertical lines, as labeled, mark the \lineClouds ~clouds.}  
\label{Abunds_rejects}
\end{figure*}

Using the molecular hydrogen columns in the first row of Table \ref{tab:NHs}, we convert molecular column densities to abundances with respect to \Htwo.
The resulting abundances are provided in Table \ref{tab:abunds} and plotted against \NHtwo~in Figures \ref{Abunds_smalls} and \ref{Abunds_rejects}. 
Abundance errors are estimated as the fractional error on the measured molecular column added in quadrature with the fractional error on the 
molecular hydrogen column for each velocity component, resulting in typical errors of 10 to 20 percent. 
The abundance results are summarized for each molecule below.

\begin{table*}
\centering
\small
\footnotesize{
\caption{Derived molecular abundances relative to \Htwo}
\label{tab:abunds}
\begin{tabular}{lrrrrrrrrrrrl}
\hline
Molecule		& \multicolumn{11}{c}{Absorption Cloud Velocity (\kmss)} 				& Units \\
			& -120	&-106	&-92	& -80	&-73	&-58	&-47	&-40	&-23	& 0 	& +20 \\
\hline
OH			& 1.8(2)& 0.69(6)&1.30(8)&1.6(1)&1.01(7)&1.3(1)	&0.8(1)	&0.67(8)&1.1(1)	&0.69(9)&1.1(2)	&$10^{-6}$\\
\cycloprop$^{[1]}$	& 2.5 	& 2.5  	& 2.5  	&2.5	& 2.5  	& 2.5  	& 2.5  	& 2.5  	& 2.5  	& 2.5 	&2.5 	&$10^{-9}$ \\
\multicolumn{2}{l}{{\it c-}H$^{13}$CCCH}
				&1.0(2)	& 1.0(3)&	&0.6(4)	& 	&  &\textless0.3&0.2(1) & 0.9(2)&1.0(2)	&$10^{-10}$ \\
\formald 		&4.1(5) &4.8(9)&1.8(2)	&4.1(5)	&3.4(2)	&2.0(2)	&2.9(3)	&7.0(7)	&2.9(4)	&5.8(16)&2.5(5)	&$10^{-7}$ \\
H$_2^{13}$CO 		&	&8.4(9)&	&4.5(9)&3.4(13)&  	&	&10(4)	&6.2(11)&10(1)	&	&$10^{-9}$ \\
SO 			& 	&1.2(1)&\textless0.3&0.6(1)&1.3(1)&0.6(2)&0.8(2)&2.5(3)	&1.1(3)	&1.4(2)	&0.9(2)	&$10^{-9}$ \\
CS 			&1.3(2)	&1.6(1)$^{[2]}$
					&0.90(6)&1.7(1)	&1.7(1)	&0.22(4)&0.58(6)&2.5(3)$^{[2]}$
											&0.72(8)&0.9(1)$^{[2]}$
														&0.46(8)&$10^{-8}$ \\
C$^{34}$S 		&	& 1.9(2)&0.3(1)	&2.0(2)	&2.6(2)&\textless1.2
									&0.6(2) &3.6(4)	&0.7(1)	&1.7(2)	&0.8(1)	&$10^{-9}$ \\
CCS 			& 	&2.8(3)	&0.6(2)	&\gsim0.8&1.9(4)&     &\gsim0.8	&3.5(4)	&1.2(3)	&1.8(2)	&1.1(2)	&$10^{-10}$ \\
HCS$^{+}$ 		& 	&2.6(4)	&1.7(6)	&	&2.4(6)	&  	&  	&1.4(4)	&\lsim2.8&2.9(4)&	&$10^{-10}$ \\
H$_2$CS 		& 	& 3.3(6)&1.0(5)	&	&$\leq$1.4&  	&1.3(1)	&5.6(8)	&\lsim2.3&1.9(4)&\lsim0.3&$10^{-10}$ \\
SiO 			& 	&\textgreater0.4&	
					&1.3(1)$^{[3]}$&1.7(1)&\textless0.4
									& 0.15(7)$^{[3]}$
										&0.37(5)&0.39(6)&1.6(2)	&1.3(2)	&$10^{-9}$ \\ 
$^{29}$SiO 		& 	&3.6(12)&	&4(3)$^{[3]}$&9(2)&  	&   	&   	&   	&12(2)	&11(2)	&$10^{-11}$ \\
$^{30}$SiO 		& 	& 2.2(9)&	&	& 	&  	&   	&   	&   	&	&	&$10^{-11}$ \\
\lneutralBrett		& 	&1.9(3)	&1.5(5)	&2.5(5)	&2.9(6)	&  	&	&\lsim6.1&4.6(11)&2.1(3)&2.2(6)	&$10^{-10}$ \\
\lBrett			& 	&4.4(3)	&	&	&11(4)	&  	&11(4) 	&   	&	&4.8(2)	&4.0(1)	&$10^{-11}$ \\

\hline
\multicolumn{13}{l}{\footnotesize{[1] The abundance of \cycloprop ~is assumed to be 2.5 $\times ~10^{-9}$.}}\\
\multicolumn{13}{l}{\footnotesize{[2] Peak optical depths are \textgreater 0.7 in CS absorption.}}\\
\multicolumn{13}{l}{\footnotesize{[3] Appears to be a wing associated with the -73 or -40 \kms ~gas instead of a separate component.}}\\

\hline
\end{tabular}}
\end{table*}
\normalsize

\subsubsection{OH}
\label{subsec:OHabund}

The (J=$\frac{3}{2}$ $\Omega$=$\frac{3}{2}$ F=1$^+$-1$^-$) and (J=$\frac{3}{2}$ $\Omega$=$\frac{3}{2}$ F=2$^+$-2$^-$) 
transitions of OH at 1665 and 1667 MHz have extremely similar line profiles and very consistent values of \Nmol ~are derived using 
the two lines, with measurements nearly always within 8 percent of one another.  In Table \ref{tab:abunds}, we report values obtained 
from a best fit to these two lines and obtain values of order $10^{-6}$.  
The (J=$\frac{3}{2}$ $\Omega$=$\frac{3}{2}$ F=1$^+$-2$^-$) line profile at 1612 MHz appears somewhat different however.  
 Column density measurements made using the 1612 MHz transition are slightly higher, by a factor of 1.5 to 3,
 than the values reported in Table \ref{tab:abunds} for the clouds at -120, -58, -46, -40, -23, 0, and +25 \kms ~clouds.
 The abundance estimate is slightly lower, by a factor of 3, in the -92 \kms ~cloud, and an order of magnitude lower than 
 the reported values at -80 and -73 \kmss.  
 OH does not appear to occupy the same volume of gas as \cycloprop, ~as evidenced by the differences in the line profiles, so
 it is not appropriate to interpret the measurements as true abundances.  However, it is noteable that while statistically significant
 scatter exists, the values of \NOH/\NHtwo ~typically decrease with \NHtwo, as is apparent from the negative slope for OH in Figure \ref{Abunds_smalls}.
 We include the values as they likely have physical significance as discussed in \S \ref{subsec:OHdisc}.

\subsubsection{CS-bearing molecules, SO, and \formald}
\label{subsec:res_CSbearing}
\begin{figure*}[!htb]
\centering
 \includegraphics[width=16.5cm]{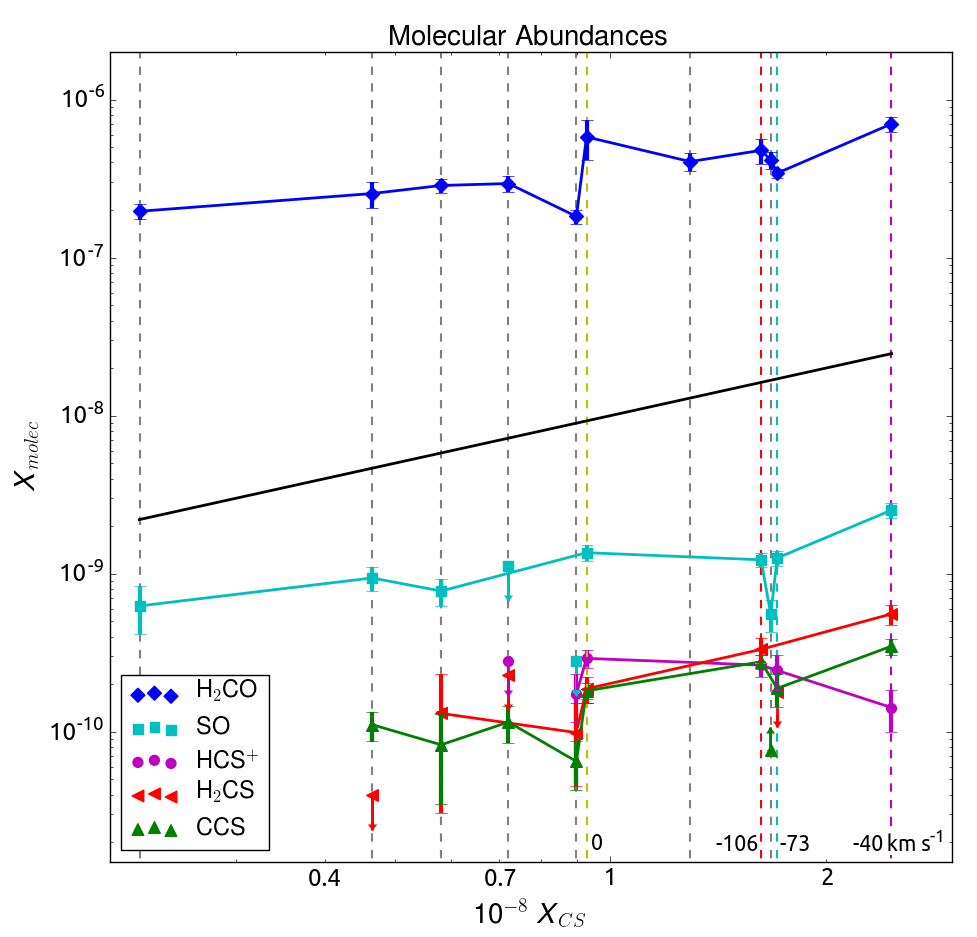}
\caption{Molecular abundances of \formald, SO, and CS-bearing species plotted against the abundance of CS.  
The black solid line represents the identity line of the CS abundance. In order from lowest to highest 
CS abundance, the -58, +20, -47, -23, -92, $\sim$0, -120, -106, -80, -73, and -40 \kms ~components are marked with vertical dashed lines. 
Colored vertical lines, as labeled, mark the \lineClouds ~clouds.}
\label{Abunds_CS}
\end{figure*}

\formald, SO, and most of the CS-bearing species share similar abundance patterns.  The CS abundance varies by an order of magnitude within the clouds, 
and no trend is evident with the cloud hydrogen column. While an abundance of $\sim$1.6 $\times~10^{-8}$ is most commonly measured, we note anomolously lower 
values of 2 to 6 $\times~10^{-9}$ in some clouds, and a particularly high value of 2.5 $\times 10^{-8}$ in the -40 \kms ~cloud.  
In addition to CS, high abundances of \formald, SO, CCS, and \HtwoCS ~are measured in the -40 \kms ~cloud.  Conversely, 
the abundance of \HCS ~is lower in the -40 \kms ~cloud than in other measured clouds.  Whereas \HCS ~is measured to have an 
abundance of $\sim$2.5 to 3~$\times~10^{-10}$ in most other clouds, it is somewhat lower at 1.4 $\times~10^{-10}$ in the -40 \kms ~cloud.  Figure \ref{Abunds_CS} shows 
the abundances of \formald, SO, CCS, \HtwoCS, and \HCS ~plotted against the abundance of CS.
The abundances of CCS and \HtwoCS ~positively correlate with the CS abundance, although scatter exists.  
On the other hand, the abundance of \HCS ~does not vary 
linearly with the CS abundance; instead it is low in the -40 \kms ~cloud and nearly constant for all other clouds. 

The abundances of \formald ~and SO also positively correlate with that of CS.  
The observed trends in $log(X_{\text{SO}})$ v. $log(X_{\text{CS}})$ and $log(X_{\text{H}_2\text{CO}})$ v. $log(X_{\text{CS}})$
are shallower than the CS-identity line, however, 
with slopes of $\sim$0.45.  Additionally, while these species show a clear correlation, statistically significant scatter is present.  

\subsubsection{SiO}
\label{subsec:res_SiO}
The abundance of SiO varies by an order of magnitude in different clouds.  Furthermore, $X_{\text{SiO}}$ does not correlate with \NHtwo~or $X_{\text{CS}}$.
In the -73 \kmss, 0 \kms, and +20 \kms ~clouds in the Galactic Center, the SiO abundance is 1.3 to 2 $\times 10^{-9}$.  
However, not all clouds believed to occur in the Galactic Bar or Galactic Center have high abundances of SiO; in
the -58 \kms ~cloud, for example, SiO is at least an order of magnitude less abundant.  In material located external to the Galactic Bar at -23 and -40 
\kmss, SiO abundances are of order 4 $\times 10^{-10}$.  

\subsubsection{\lneutralBrett ~and \lBrett}
\label{subsec:res_lC3H}
The abundances of \lneutralBrett ~range from $\sim$1.5 to $\sim$5 $\times 10^{-10}$, with a median value of 2.2 $\times 10^{-10}$, whereas the 
abundance of \lBrett ~varies from  4 $\times 10^{-11}$ to 11 $\times 10^{-10}$.  
The \lneutralBrett ~and \lBrett ~abundances do not appear to correlate with \NHtwo ~or with \XCS.

\subsubsection{Isotopologue Ratios}

\begin{figure*}[!h]
\centering
 \includegraphics[width=16.5cm]{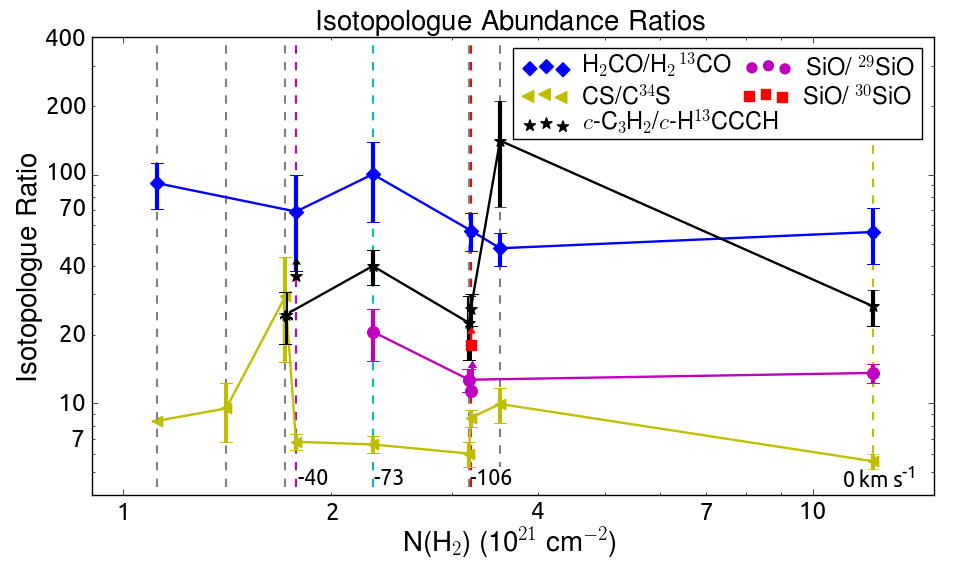}
\caption{Isotopologue abundance ratios.  In order from lowest to highest 
hydrogen column, the -80, -47, -92, -40, -73, +20, -106, -23, and $\sim$0 \kms ~components are marked with vertical dashed lines.
Colored vertical lines, as labeled, mark the \lineClouds ~clouds.}
\label{Abunds_Isos}
\end{figure*}

The abundance ratios (or equivalently, integrated column density ratios) between isotopologues show significant variation 
within different clouds (Table \ref{tab:isotopologues} and Figure \ref{Abunds_Isos}).  
The $^{12}$C/$^{13}$C ratio is probed by \formald ~and \cycloprop.  Values of \formald/\formaldIso
~range from 45 to 70 for the -106, -40, -23, and \zero \kms ~clouds, 
while higher values of $\sim$95 are found in the material at -73 and -80 \kmss.  
Thus, the ratios of \formald/\formaldIso ~do not show any obvious pattern with Galactocentric distance.
Values of \cycloprop/\cyclopropIso ~are $\sim$25 in the Galactic Center clouds at 
-106, -92, \zero, and +20 \kmss, slightly higher at $\sim$40 in the -73 \kms ~cloud located deep within the Galactic Bar, 
and significantly higher, at \gsim 100 \kms ~in material in the Galactic disk at -23 \kmss. 

\begin{table*}[!htb]
\small
\caption{Isotopologue abundance ratios}
\label{tab:isotopologues}
\begin{tabular}{lrrrrrrrrrr}
\hline
Molecules    		& \multicolumn{10}{c}{Absorption Cloud Velocity (\kmss)} 		\\
							&-106	&-92	& -80	&-73	&-47	&-40	&-23	& 0 	& +20 	\\
\hline
{\it X}(\cycloprop)/{\it X}({\it c-}H$^{13}$CCCH)	& 26(4)	& 24(6)	&	& 40(7)	&  	&\textgreater36 
													&141(70)&27(5) 	&23(7)	\\
{\it X}(\formald)/{\it X}(H$_2^{13}$CO)			& 57(10)& 	&91(21)	&100(38)&	&69(31)	&48(8) 	&56(15)	&	\\
{\it X}(CS)/{\it X}(C$^{34}$S)				& 8.7(8)$^{[1]}$
							&29(14)	&8.4(1)	&6.6(6)$^{[1]}$	&10(3)&6.8(5)$^{[1]}$  	
													&10(2)	&5.6(4)$^{[1]}$ & 6.0(7)	\\
{\it X}(SiO)/{\it X}($^{29}$SiO)			&\gsim11&	&	&21(5)$^{[2]}$	
											&   	&   	&	&13(1)  & 13(2)	\\
{\it X}(SiO)/{\it X}($^{30}$SiO)			&\gsim18&	&	& 	&   	&   	&  	& 	&	\\
{\it X}($^{29}$SiO)/{\it X}($^{30}$SiO)			& 1.6(2)&	&	& 	&   	&   	&   	& 	&	\\
\hline
\multicolumn{11}{l}{[1] Peak CS absorption has a high optical depth of $\tau_0$ \textgreater~0.7.}\\
\multicolumn{11}{l}{[2] Absorption in the velocity range of -85 to -67 \kms ~was included in this measurement.}\\
\hline
\end{tabular}
\end{table*}

Isotope ratios of silicon are probed by measurements of $^{28}$Si, $^{29}$Si, and $^{30}$Si.
In the \zero and +25 \kms ~material, we obtain $^{28}$SiO/$^{29}$SiO = 13, and in the -73 \kms ~cloud, 
we obtain a higher value of 21. Due to the presence of an interloping line near -106 \kms ~in the spectrum of $^{28}$SiO, 
we are unable to measure the $^{28}$SiO/$^{29}$SiO ratio in that cloud.
The \twentyNineThirty ~ratio is measured to be 1.6 in the -106 \kms ~cloud.

The \thirtyTwoThirtyFour ~ratio is probed by  CS/C$^{34}$S.  For all clouds except for the -92 \kms ~cloud, the measured isotopologue ratio 
ranges from 5 to 10.  No trend is evident with hydrogen column or with Galactocentric distance.  Within the components 
with high optical depths of peak $\tau$ \textgreater~0.7 in CS absorption, 
in which the measured isotopologue ratio may be depressed from the true value,
we obtain measurements ranging from 5.6 to 8.7.  In components with lower optical 
depths (of peak $\tau$ \textless 0.5) in the CS (1-0) line, values of 6 to $\sim$10 are observed.  
In the -92 \kms ~component, we obtain CS/C$^{34}$S = 29 $\pm$~14, and it is unclear why this component is anomolous.

\section{Cloud Hydrogen Columns:  Considerations for Interpreting Abundances}
\label{sec:Hydrogen}

We have obtained new measurements of the molecular hydrogen columns within each cloud in the line-of-sight to Sgr B2, and of abundances 
of eleven molecules and their isotopologues.  Before discussing the implications of these measurements, we 
explore the validity of using \cycloprop ~as a proxy for \Htwo~and other preliminaries required for intrepreting the abundance
measurements.

\subsection{Molecular Hydrogen Columns by Conversion from \cycloprop}
\label{subsec:NH2_cC3H2}

In order to maximize the self consistency in our measurements, we probe the molecular hydrogen column 
using a molecule that is observed within our dataset; we select 
\cycloprop ~as the best proxy for \Htwo ~observable at centimeter wavelengths.
In millimeter wavelength studies of diffuse and translucent clouds, hydrogen columns have typically been estimated  
by assuming a constant abundance of either CO or \HCO ~\citep[e.g.][]{Irvine87,GN96}.  
To the best of our knowledge, molecular hydrogen columns within 
diffuse and translucent clouds have not previously been estimated by conversion from \Ncycloprop.
However, we suggest that the assumption of a constant abundance of \cycloprop ~is reasonable and in fact may provide 
a more direct and accurate method of estimating the molecular hydrogen column than conversion from CO or \HCO.

Based on observations from radio to UV frequencies, the hydrocarbon molecules CH, CCH, and \cycloprop 
~appear to be constant in abundance with respect to one another and with respect to \Htwo ~in diffuse and translucent gas.  
The constancy of \NCH/\NHtwo ~over the range of 19~\textless~log(\NHtwo)~\textless~22 has been robustly demonstrated and directly 
calibrated by \citet{Sheffer2008} in an analysis of 90 sightlines with CH absorption and \Htwo ~Lyman band observations.  
This hydrogen column range completely spans the domain of interest, from diffuse to the transition to dark molecular gas.
\NCCH/\NCH ~has also been shown to be constant in an analysis of Herschel HIFI data towards three sightlines that each contain multiple 
distinct diffuse and translucent clouds \citep{Gerin2010b}.  
The very tight correlation of $\tau_\text{CCH}$~v. $\tau_\text{CH}$ 
was demonstrated over hundreds of channels in each sightline.  Finally, multiple studies substantiate the constancy of \Ncycloprop/\NCCH 
~in diffuse and translucent gas based on analyses of 26 distinct clouds in 9 sightlines \citep{LL2000, Gerin2011, LisztcC3H2}.  
In these sightlines, CCH and \cycloprop ~spectra show remarkable correlation, with a consistent opacity ratio
present in channels from negligible optical depths to peak values \citep{LL2000,Gerin2011}, and at various Galactocentric distances 
from the Galactic Center to the outer disk \citep{Gerin2011}. 
In addition to having a stable abundance in diffuse and translucent clouds, \cycloprop ~is observed to have precisely the same 
abundance with respect to molecular hydrogen in the dense ($n \sim 10^5$~\cmt) Horsehead PDR \citep{Pety05}, indicating that the  
value of N(\cycloprop)/N(\Htwo) may be highly stable in UV-irradiated gas over a wide range of temperatures and densities.  
Therefore, the hydrocarbon species CH, CCH, and \cycloprop ~appear to 
be good tracers of \NHtwo, as their abundances do not systematically vary with \NHtwo ~over the column range of diffuse and translucent clouds.

On the other hand,
the assumption of a constant CO abundance is easily contested in diffuse and translucent gas, given a growing body of evidence 
suggesting the presence of CO-dark molecular gas in UV-irradiated environments \citep[see e.g.][]{COdark}, and evidence that $N_{\text{CO}}$/\NHtwo
~varies by a factor of \textgreater 100 in diffuse and translucent clouds \citep{Sonnentrucker07}.  
The work of \citet{Sheffer2008} further demonstrates the variability in the CO abundance in diffuse and translucent clouds, 
with a sample of more than 100 sightlines with direct measurements of \NHtwo ~and \NCO; 
the data reveal that \NCO/\NHtwo ~systematically varies with \NHtwo, with a steep slope of 
log(\NCO)/log(\NHtwo) = 3.1(7) above \NHtwo = 20.4.  Further, his data reveal substantial scatter, by nearly two orders of magnitude for a given 
hydrogen column.  CO is thus a very poor indicator of total gas column in diffuse and translucent clouds.


While \HCO ~abundance variation does not appear to be as substantial as for CO, \HCO:CH and \HCO:CCH ~have been shown to 
have a non-linear dependence on \NHtwo ~\citep{LL2000, Gerin2010b}.
Like CO, if \NCH/\NHtwo ~is adopted as constant, then the abundance of \HCO ~increases in higher column density systems.  

Finally, conversions from CO and \HCO are further complicated by the fact that their rotational transitions tend to be optically thick.
While the abundances of the \twelveC ~isotopologues are assumed to be constant for the purpose of conversion to \NHtwo, the column densities of the \twelveC
~isotopologues often must first be estimated by conversion from $^{13}$CO or H$^{13}$CO$^{+}$.  Previous authors have 
assumed differing ratios of $^{12}$C/$^{13}$C isotopologues internal to the Galactic Center, Bar and disk.  
This is dangerous for numerous reasons, including that some \los ~clouds have poorly constrained 
Galactocentric distances.   
As is clear from our observations of isotopologues (Figure \ref{Abunds_Isos}), significant variation occurs in the
isotopologue ratios in different clouds, and for many molecules, these cannot be accounted for by the 
Galactocentric distances presumed.  It is also well established that isotope fractionation occurs creating discrepancies between 
the true isotope ratios (i.e. \twelveThirteen) and the isotopologue ratios (i.e. \twelveCOThirteenCO ~and \twelveHCOThirteenHCO).

Conversion from \cycloprop ~is more direct, as it utilizes optically thin transitions of a molecule that has been systematically explored 
in diffuse and translucent clouds over a broad range of cloud thicknesses and throughout the 
Galaxy \citep{LL2000, Gerin2011, LisztcC3H2} and does not require an isotope ratio conversion.  
As such, \cycloprop ~appears to be the best means of determining the hydrogen column at our disposal, 
and we recommend further exploration and precise calibration of the \cycloprop ~abundance via observations and modeling.
Despite the various issues with using CO and \HCO, we note general agreement between our measured values of \NHtwo ~and previous measurements 
of the same clouds in the \los ~to Sgr B2(M) located $\sim$45 arcsec South (Table \ref{tab:NHs}).

\subsection{Defining Cloud Borders}
\label{subsec:divisions}

We also point out differences in how we distinguish between clouds.  In the previous studies, authors have combined multiple cloud components
for hydrogen column and molecular abundance estimates.  For example, the clouds at -47 and -40 \kms ~were treated as a single cloud 
in analyses by previous authors \citep{Irvine87,GN96, WirstromNH3}. As the -47 and -40 \kms ~clouds have demonstrable differences in 
molecular abundances, for example in the abundance of CS, we treat these systems as distinct.  For the purpose of follow-up studies
of the chemistry within these environments, we consider it prudent to treat the clouds as specifically as is possible rather than present 
average abundance values which may not be representative of any environment in the cloud.  
In this vein, we might consider dividing the clouds further.  For example, the -40 \kms ~cloud contains two narrow velocity components
that are clearly resolvable in CS absorption.  However, for many of the lower \StoN ~transitions, 
we do not have the \StoN ~required to distinguish the two components.  In this manner, we have maximized the specificity 
with which we may treat individual clouds.  

Finally, it is standard to interpret a hydrogen column as an extinction value by Equation \ref{eq:AVestimates}.
Obviously, the implied extinction values will vary depending on where 
the clouds are distinguished.  For example, if the -47 and -40 \kms ~clouds are adjacent and do not contain UV-radiation sources 
located between the two clouds, then it is, indeed, physically appropriate to treat the 
two clouds as a combined system with a higher extinction.  
However, we know little about the three-dimensional geometry of the clouds and incident radiative fields.
In fact, substantial differences may occur in the incident
UV radiation field depending on the Galactocentric distance and on the precise local environment of the cloud, so that 
two clouds with similar hydrogen columns may experience very different radiative fields at their centers. Secondly, 
cosmic rays, known to be multiple orders of magnitude more prevalent in diffuse clouds the Galactic Center \citep{lePetitCRio}, 
may induce local UV-radiation fields very deep within the cloud, and X-rays which are 
particularly prevalent in the Galactic Center can penetrate the diffuse and translucent clouds 
\citep{AmoB09}. Finally, the geometry and structure of the diffuse and translucent ISM
can significantly affect the radiative field experienced within the cloud.  
Thus, we do not put much emphasis on the total hydrogen column as a determinant of the chemistry observed within the clouds, and indeed
we do not observe significant trends with total hydrogen column in Figures \ref{Abunds_smalls} and \ref{Abunds_rejects}.  
Instead, we point out differences between clouds located internal and external to the Galactic Center and Galactic Bar, and look for 
clues to the structure of different clouds.  

Translucent clouds have typical central extinctions of 1 \textless~\AVo~\textless~2.5.  
Many of the clouds, as divided in this work, do not meet this threshold, having central extinction values of 0.5 to 1.0 
(Table \ref{tab:NHs}). This clearly depends on how the clouds are binned, however we point out that many of the clouds 
considered may be intermediate between diffuse clouds and translucent clouds as defined by this extinction threshold, 
and are likely transitional regions.  
This result deserves further exploration however, as the average molecular hydrogen fractions, estimated at 0.65 to 0.95, 
are significantly higher than the values of 0.4 to 0.5 commonly observed in diffuse gas \citep{LisztcC3H2, Snow&McCall}.  
Nonetheless it may be reasonable to consider many of the these systems as diffuse or transitional clouds based on the dominant 
form of carbon in the material.
The classification system for diffuse atomic, translucent, and dark clouds introduced by \citet{Snow&McCall} defines translucent clouds based on the dominant form of carbon, 
as having $f(\text{C}^+)$~\textless~0.5 and $f(\text{CO})$~\textless~0.9.  Studies of {\sc OI}, {\sc CII}, and $^{13}$CO line profiles 
towards Sgr B2(N) show that carbon is predominantly in the neutral and ionized forms within this material \citep{Vastel2002}, 
indicating that it may indeed be transitional between diffuse molecular and translucent.

\section{Discussion}  
\label{sec:discussion}

We now discuss the implications of the results described in \S \ref{sec:PRIMOSresults} for the locations and physical structure of the line-of-sight clouds.

\subsection{Discussion of Isotopologue Ratios and Fractionation }  
\label{subsec:disc_isos}

The observed isotopologue ratios of \formald ~and \cycloprop~suggest systematic differences between the two.  
Therefore, it is clear that isotope fractionation 
is present for at least one of the species.  While isotope ratios of \cycloprop ~have not been studied closely, models by \citet{Langer84} 
indicate fractionation occurs for the \formald/\formaldIso ~ratio, such that this provides an upper limit to \twelveThirteen, while the \twelveCOThirteenCO
~ratio provides a lower limit. Extensive studies of the \twelveThirteen ~ratio 
have been published using isotopologues of CO, \formald, CN, and CH$^{+}$ \citep{WilsonIsos,WilsonRood94,StahlGW08,Milam05}.
While the studies show definitive evidence for variation in different sources of the same galactocentric distance, 
a general trend of increasing \twelveThirteen ~ratios with increasing Galactocentric distance $D_\text{GC}$ is well established, with typical
values of $\sim$20 in the Galactic Center, $\sim$50 at a Galactocentric distance of 4 kpc, and $\sim$70 at the solar Galactocentric distance.   
In most clouds located internal to the Galactic bar, the \formald/\formaldIso ~ratios measured are a factor of two to three
higher than expected, and the reported values do not show a clear trend with Galactocentric distance, but do show statistically significant variation.  
The ratios measured by \cycloprop/\cyclopropIso, namely $\sim$25 in the Galactic Center, 40 at \lsim 1 kpc, \textgreater36 at 3 kpc, and \gsim100 at 4 kpc, 
show a very clear trend with Galactocentric distance, and are very similar to the previous measurements of \twelveThirteen ~ratios. 

The isotope ratio of \thirtyTwoThirtyFour ~has also been investigated \citep{Frerking80, SulfurIsos_Chin96, Mauersberger96}, 
but it is not as thoroughly researched as the carbon isotope ratio, and disagreement persists over 
what trends, if any, are present with Galactocentric distance.  Previous studies typically obtained values of 24.4~$\pm$~5, whereas we observe substantially 
lower values of 5~\textless~CS/C$^{34}$S~\textless~10 for all clouds with the exception of the -92 \kms ~cloud, 
which has a higher value of 29 $\pm$ 15.  The measured ratios do not show any apparent trend with Galactocentric distance.  
Generally, chemical fractionation is not believed to affect sulfur chemistry substantially, 
as the zero point energies of $^{32}$S and $^{34}$S are quite similar.  

Finally, the ratios of \twentyEightTwentyNine ~and \twentyNineThirty ~have not been systematically investigated in the Galaxy.  
In the \zero and +25 \kms ~material, we obtain $^{28}$SiO/$^{29}$SiO ~=~13, and in the -73 \kms ~cloud, 
we obtain a higher value of 21 $\pm$~5.  Additionally, we measure $^{29}$SiO/$^{30}$SiO~=~1.6 $\pm$~2.  Within the solar system, a value of 
\twentyEightTwentyNine ~=~19.7 is observed \citep{WilsonIsos,  Penzias81}. 
Assuming that chemical fractionation does not significantly affect the SiO isotopoluges, our data indicates that 
the \twentyEightTwentyNine ~in the -73 \kms ~cloud is consistent with the local value, while $^{29}$Si is enhanced relative 
to the main isotope in the \zero ~and +25 \kms ~material located in the Galactic Center.  Additionally, early investigations 
of \twentyNineThirty ~have determined values of 1.5 without a gradient with galactocentric distance \citep{Penzias81, HuttemeisterSiO}. 
The measured value of \twentyNineThirty ~in the -106 \kms ~gas is consistent with this, supporting the supposition that 
the value of \twentyNineThirty ~does not systematically vary with Galactocentric distance.  

Given these observations, a few trends can be pointed out.  First, \formald/\formaldIso ~is not a good probe of the \twelveThirteen ~ratio, 
and includes significant fractionation.  If the fractionation can be understood, this may be interesting from a chemistry perspective.  
Secondly, the values of \cycloprop/\cyclopropIso ~exhibit a trend with Galactocentric distance, and are consistent with previous measurements 
of \twelveThirteen ~ratios.  This suggests that isotope fractionation is not significant for these two isotopologues of \cycloprop, 
so that \cycloprop ~can probe the \twelveThirteen ~ratio accurately, with the added advantage over CO and HCO$^+$ of being optically thin.  
Third, $^{34}$S is substantially enhanced in the clouds measured here compared to previous measurements in other clouds, assuming that 
CS/C$^{34}$S does not undergo fractionation. The PRIMOS data support the supposition that there is no gradient in \thirtyTwoThirtyFour ~with 
Galactocentric distance.  

Finally, the isotopologue ratios provide clues as to the locations of the \los ~clouds in the Galaxy.  For example, the 
-73 \kms ~cloud has values of \cycloprop/\cyclopropIso ~that are intermediate between those observed in Galactic Center 
clouds and those of clouds located in to the Galactic disk and previously measured locally; further, the $^{28}$SiO/$^{29}$SiO ~ratio in the -73 \kms ~cloud 
is clearly distinct from the values measured in the Galactic Center clouds and is comparable to the value obtained in the solar system.
These observations are consistent with the -73 \kms ~material being located external to the Galactic Center and in the Bar.  The observations therefore 
lend further support to the approximate locations of the clouds in the Galaxy.

\subsection{Sulfur-bearing Chemistry: CS, SO, \HCS, CCS, and \HtwoCS}
\label{subsec:disc_CS}

Clearly, diverse sulfur-bearing chemistry is observed within the diffuse and translucent clouds.  
The abundances of a few of the sulfur-bearing molecules detected here have been 
investigated in previous studies of diffuse and translucent clouds, most recently by \citet{LL_CS}
and \citet{NeufeldCS}.  \citet{LL_CS} measured the abundances of CS, SO, and HCS$^{+}$ in diffuse clouds 
absorbing against extragalactic point sources at high Galactic latitude, while \citet{NeufeldCS} 
measured the abundances of CS and SO in diffuse and translucent clouds in the Galactic plane.  
In this section, we discuss our observational results in frequent reference to these two papers. 
Additionally, \citet{Drdla} reported abundance measurements of CS and \HCS ~in diffuse and/or translucent 
clouds at high Galactic latitude, and contributed the most comprehensive theoretical model predictions for 
sulfur-bearing molecule abundances in diffuse clouds published to date.  

In the following discussion, we come to the following main conclusions on sulfur-bearing chemistry based on the data shown here and 
in the context of the three forementioned papers.
 First, in clouds in the Galactic disk, the data suggests that CS, and likely SO to a lesser degree, preferentially 
inhabit the highest density regions of the diffuse and translucent clouds, with abundance enhancements in high density material.  
 Secondly, whereas the abundances of CS and \HCS ~are predicted to be closesly coupled \citep{Drdla, LL_CS}, we do not observe this
and our abundance ratios of CS:\HCS ~disagree with the observations of \citet{LL_CS} and \citet{NeufeldCS}.  
 Third, SO and \HtwoCS ~are severely overabundant as compared to model predictions \citep{Drdla}, 
and model predictions for CCS in diffuse and translucent clouds have not been published.  With the wealth of data
published, and with recent advances in chemical models, we suggest the sulfur-bearing chemistry in diffuse clouds should
be revisited.

\subsubsection{CS: Abundances and Implications for Cloud Structure}

Our reported CS abundances vary by more than an order of magnitude, with a few clouds having values in the range of 2 to $7 \times 10^{-9}$ 
and multiple clouds in the range of 1 to $2.5 \times 10^{-8}$.  The latter set of clouds have high abundances of CS 
as compared to most interstellar environments, with values similar to the abundances 
observed in the hot core in Sgr B2(N) and in the Orion Molecular Ridge \citep{HEXOSsgr, HEXOSorion} 
and a factor of two to three times higher than in the dense Horsehead PDR \citep{Goi_CS} and the dark cloud TMC-1 \citep{OhishiCS}.
The measurements presented here are in approximate agreement with values reported by \citet{GN96} in the \los ~to Sgr B2(M), of 2 to $5 \times 10^{-8}$. 
However, the CS abundances measured in most of the clouds in the sightlines to Sgr B2(N) and (M) 
are significantly larger than those reported in diffuse clouds towards other sightlines by \citet{LL_CS} and \citet{NeufeldCS}.  
In both latter papers, the authors derived abundances that typically ranged 
from 1 to 5 $\times 10^{-9}$, although one of the ten clouds measured by \citet{NeufeldCS} was larger, at $10^{-8}$. 
While \citet{LL_CS} typically obtained lower values than we measure, they noted that the CS abundance appears sporadic 
in the diffuse clouds; in some clouds with significant column densities of \HCO, and presumably therefore of \Htwo, CS was undetected, 
whereas it was clearly detected in other clouds with lower \HCO ~columns.  In the data presented here, by 
\citet{LL_CS}, and by \citet{NeufeldCS}, the abundance of CS does not appear to scale with the total molecular 
or neutral hydrogen column.  Additionally, we do not observe systematically higher or lower abundances in the 
Galactic Center or Bar clouds compared to disk clouds.


While there is no clear differentiation between Galactic Center/Bar clouds and disk clouds based on 
the CS abundance values observed, there is an apparent distinction between the shapes of the CS profiles in Galactic Center clouds compared to disk clouds.  
As described in \S \ref{subsec:profiles} the CS absorption in Galactic Center and Bar clouds, for example at -106, -73, and 0 \kmss, includes 
absorption over the same moderately broad velocity ranges as do \cycloprop ~and \formald.  On the other hand,
the profile of CS contains very sharp, narrow peaks in the clouds located external to the Galactic Bar, namely in 
the -47 and -40 \kms ~clouds believed to be located in the 3-kpc arm at the outer edge of the Bar and in the -23 \kms ~cloud. 
This suggests that CS is highly sensitive to the physical conditions within the clouds, and may indicate a systematic difference 
in the structure of clouds that are located in the Galactic Center compared to those present in the disk. 

In the Galactic disk clouds, in which CS contains narrow peaks with little broad component absorption as compared to 
\cycloprop ~and \formald, we suggest that CS is significantly more abundant in the highest density regions of the clouds.  
While chemical models notoriously underpredict abundances of sulfur-bearing species in the diffuse ISM,
the known reactions can achieve higher CS abundances at higher physical densities and/or higher visual extinctions
\citep[see e.g.][and references therein]{LL_CS, NeufeldCS}. 
However, as the total hydrogen column has been shown to be a poor predictor of the CS column density and abundance 
in the data presented here, in \citet{LL_CS}, and in \citet{NeufeldCS}, it appears that the CS abundance is more sensitive 
to density conditions than to extinction in the diffuse and translucent medium.  This claim is also supported by previous observations; 
the work of \citet{Greaves92} suggests that CS resides in material that is significantly more dense than average
in the line-of-sight to Sgr B2(M).  While the mean densities in the 
clouds appear to be $n \sim$~300 to 500 \cmt, CS resides in material with densities of
4000 \lsim~$n$~\lsim~15\,000~\cmt~as estimated by a Large Velocity Gradient (LVG) analysis of CS (1-0) and (2-1). 

If CS is sensitive to density but not directly correlated with extinction, then this suggests that the density profile in the diffuse clouds 
is not a simple, smooth function of extinction.   
The CS profiles therefore suggest that the diffuse and translucent clouds in the disk have a clumpy physical structure, as the 
CS observations cannot be accounted for by a gradual density enhancement with \AV.  Furthermore, the differences between 
the CS profile and the profiles of most other molecules suggests that CS is preferentially residing in a different chemical environments. 
This in turn suggests a layered chemical structure, with a greater number 
of chemical environments than previously described.

\subsubsection{SO}
The absorption profile of SO suggests that it too has an enhanced abundance in the 
highest density regions in the -47, -40, and -23 \kms ~clouds, with narrow peaks dominating the absorption 
in the -47 and -40 \kms ~clouds especially.  The abundance of SO shows a positive linear correlation 
with the CS abundance but does not vary as much as CS does (Figure \ref{Abunds_CS}).  Therefore the SO abundance may be enhanced, but by a lower 
factor than CS, in the highest density regions. 
\citet{LL_CS} also found a positive but loose correlation between SO and CS in diffuse gas. 

The SO abundances, with a median value of 1.1 $\times 10^{-9}$, are similar to those measured by 
\citet{LL_CS} and \citet{NeufeldCS}.  \citet{LL_CS} notes that gas phase models of sulfur-bearing chemistry 
underpredict this value by multiple orders of magnitude.  Even upon considering shocks and turbulent dissipation, 
models underpredict the SO abundance by an order of magnitude \citep{NeufeldCS}.
Because the abundances of CS are significantly higher in most clouds along the sightline to Sgr B2(N),
we obtain significantly larger values of the ratio \NCS/\NSO ~than do \citet{LL_CS} and \citet{NeufeldCS}.  
In most clouds, these authors found that \NCS/\NSO ~= 2, although a higher value of 5.8 was present in the cloud 
with the anomolously high CS abundance, mentioned above, in \citet{NeufeldCS}.  We obtain values of \NCS/\NSO 
~ranging from 3 to $\sim$30, with a median value of 9.  In other Galactic interstellar environments, ratios of \NCS/\NSO ~have been observed to vary significantly in 
different clouds; even in clouds in similar evolutionary states and with apparently similar conditions, very different values 
may be observed \citep[e.g.][]{Gerin97}.

\subsubsection{\HCS, CCS, and \HtwoCS}
Among the lower abundance CS-bearing species,
the results demonstrate a positive linear correlation between 
the abundances of CCS and \HtwoCS ~with CS (Figure \ref{Abunds_CS}), although scatter exists, 
due in part to baseline instability and line confusion in the lower signal transitions. 
On the other hand, \HCS ~displays a different abundance pattern, with a fairly consistent abundance 
in all clouds except for the -40 \kms ~cloud.  While observation of additional transitions of \HCS 
~should be conducted to confirm this trend, the depressed abundance of \HCS ~in the -40 \kms ~cloud is notable 
because all other sulfur-bearing species have higher abundances in this cloud than in any other.  
If a larger fraction of the -40 \kms ~cloud consists of higher density material, 
it may be that the \HCS ~abundance is lower in the highest density regions of diffuse and translucent clouds.
The abundance patterns thus suggest that the ratios of \NHCS/\NCS, \NHCS/\NCCS, and \NHCS/\NHtwoCS
~appear to be sensitive to cloud conditions within diffuse and translucent clouds, making these ratios potentially useful probes of physical 
conditions.  Although in a distinct environment, \citet{Corby15} observed that CS, CCS, and \HtwoCS 
~have indistinguishable spatial distributions in Sgr B2,
while \HCS ~has a distinct spatial distribution, further supporting the 
trend that CS, CCS, and \HtwoCS ~track each other, and that the ratios of \HCS ~to other CS-bearing species are sensitive to 
physical conditions. 

According to chemical models, the molecules \HCS ~and CS are among the most closely related species present in the ISM, 
as they participate in a direct exchange in both the formation and destruction of CS \citep{Drdla, LL_CS}.  
In the reaction network considered by \citet{Drdla} and \citet{LL_CS},
CS is believed to be formed primarily from the dissociative recombination reaction of 
\begin{equation} \label{eq:HCSdisRecomb} 
\text{HCS}^+ + e^- \rightarrow \text{CS} + \text{H}, 
\end{equation}  
where \HCS ~is first formed by reactions beginning with S$^{+}$.
However, recent experiments indicate that this product channel occurs in only 19 percent of collisions, with 81 percent of collisions forming CH + S \citep{MontaigneCS}.
Three mechanisms dominate the destruction of CS in these networks.  
In the first two, 
photoionization and ion-exchange reactions destroy CS to form CS$^+$, by 
\begin{equation} \label{eq:CSphotoIo} 
\text{CS} + \gamma \rightarrow \text{CS}^+ + e^-
\end{equation} 
and
\begin{equation} \label{eq:CSionExchange} 
\text{CS} + X^+ \rightarrow \text{CS}^+ + X,
\end{equation} 
where $X^+$ is a cationic species and $X$ is the corresponding neutral species. 
CS$^+$ then quickly reacts to form \HCS:
\begin{equation} \label{eq:CSio_recomb} 
\text{CS}^+ + \text{H}_2 \rightarrow \text{HCS}^+ + \text{H}.
\end{equation}  
In the third destruction route, CS reacts with H$_3^+$ to form \HCS ~directly: 
\begin{equation} \label{eq:CSandH3plus} 
\text{CS} + \text{H}_3^+ \rightarrow \text{HCS}^+ + \text{H}_2.
\end{equation}  
As CS is formed by a reaction of \HCS ~and the dominant destruction pathways for CS form \HCS,
we would expect the abundances of CS and \HCS ~to be delicately balanced, and for 
the two abundances to track each other.  While chemical models of sulfur-bearing chemistry in diffuse clouds  
underpredict the abundance of \HCS ~by multiple orders of magnitude, \citet{LL_CS} determined that if the observed abundance of \HCS ~is 
injected into a diffuse cloud, it is possible to account for the CS abundances observed in their sample, further 
emphasizing the theoretical importance of \NCS/\NHCS. In clouds in the Galactic Center, where 
\Hthreeplus ~is \gsim\,10 times more abundant compared to diffuse clouds in the disk \citep{Oka05}, 
we might expect an offset in the ratio, tending towards a higher relative abundance of \HCS.

In stark contrast to this expectation, the abundance of \HCS ~appears to be uncorrelated with that of CS.  \HCS ~exhibits a typical 
abundance of 2.5 to $3 \times 10^{-10}$ in this study, and this value closely matches the measurements made in the diffuse clouds sampled by \citet{LL_CS}.  
On the other hand, we observe very large variations in CS abundances, by a factor of \gsim\,10 in our study and \gsim\,25 if 
we also consider the measurements made by \citet{LL_CS} and \citet{NeufeldCS}.  Furthermore, instead of observing more \HCS ~relative to 
CS in the Galactic Center clouds due to the higher abundance of \Hthreeplus, we see the opposite.  Whereas \citet{LL_CS} measured $N_{\text{CS}}$/$N_{\text{HCS}^+}$ 
= 13.3 $\pm$ 1.0 in diffuse clouds in the Galactic disk, we find much higher values (and therefore lower \HCS ~abundances relative to CS)
ranging from 50 to 70 in most Galactic Center/Bar clouds. 
The constancy of the \HCS ~abundance in clouds measured in this study and in \citet{LL_CS} indicates that 
the \HCS ~abundance is (1) consistent over a wide range of densities and extinction values, (2) independent of the CS abundance, and (3) independent 
of the Galactocentric distance.  As the same values are measured in absorbing clouds in the Galactic disk as in the Galactic Center and Bar, 
it appears that the \HCS ~abundance is not highly sensitive to the CR-ionization rate or X-ray fluxes in diffuse gas.

CCS and \HtwoCS ~have similar abundances to \HCS, with typical values of 1 to 3~$\times 10^{-10}$; like \HCS, CCS and \HtwoCS ~are significantly 
more abundant than predicted by models appropriate in diffuse, translucent, and/or any other UV-irradiated material \citep[e.g.][]{Drdla, Goi_CS}.
In fact, the observed enhancement of \HtwoCS ~is five orders of magnitude larger than predicted by \citet{Drdla}, and no prediction was published for CCS.
With the recent observational constraints on sulfur-bearing chemistry in diffuse and translucent clouds provided 
here, in \citet{LL_CS} and \citet{NeufeldCS}, and in dense PDRs \citep{Goi_CS}, a revised theoretical treatment of sulfur-bearing chemistry, 
and particularly of CS-bearing chemistry, is warranted.  
More advanced chemical models including PDR and gas-grain models with updated reaction rates for sulfur-bearing species are being developed; 
the application of these to physical conditions appropriate in diffuse and translucent clouds could prove effective for reproducing the observed chemistry.
More importantly, a thorough theoretical treatment 
of the sulfur-bearing chemistry may illuminate the patterns embedded in the varying abundance ratios, producing a more nuanced 
understanding of what the chemistry tells us about the physical and radiative structure of the diffuse and translucent ISM. 
For example, the ratio of \NCS ~to \NHCS ~may prove an excellent probe of physical conditions, however a solid theoretical understanding 
of the reaction networks is requisite for interpreting the ratios.  

\subsection{Oxygen-bearing Chemistry: OH, \formald, and SiO}
\label{subsec:OBearing}
\subsubsection{Interpreting OH Absorption}
\label{subsec:OHdisc}

OH has four hyperfine transitions in the ($^2\Pi_{3/2} ~J = \frac{3}{2}$)  
state, at 1612, 1665, 1667, and 1720 MHz.
All four transitions can exhibit masing, typically in high density (10$^6 \leq n $~\lsim$~10^9$ \cmt) environments \citep{Gray91,Elitzur92}
associated with star formation, shocks, and AGB winds.  Under lower density conditions in which strong masing is not produced, 
thermal absorption or emission line profiles are observed.  However, to the best of our knowledge, 
there have been no reported observations in which the four hyperfine transitions are in LTE.  Instead, recent work 
by \citet{Ebisawa_OH} has demonstrated that non-LTE excitation of the ground state hyperfine transitions of OH 
are ubiquitous in clouds that do not host strong masing, including a translucent cloud, 
a cold dark cloud, and a PDR, and additional reports of similar behaviors have been reported commonly \citep[see references in][]{Ebisawa_OH},  
although the reported instances tend to involve emission line sources. In the study conducted by \citet{Ebisawa_OH}, the 1665 and 1667 MHz 
transitions produce emission that is more consistent with LTE conditions over a wider range of physical conditions, while
the 1612 MHz component appears in absorption against the CMB background and enhanced emission is 
present in the 1720 MHz line. The divergence from LTE occurs by a collisional excitation mechanism, 
even at extremely low physical densities of $\sim$10 \cmt ~\citep{Elitzur92}.  

In \S\ref{subsec:OHabund}, we described that the 1665 MHz and 1667 MHz transitions have very consistent absorption profiles, 
with primarily broad (\delv~$\geq$~10~\kmss) absorption components, resulting in statistically identical OH column density measurements.
However, the profile of the 1612 MHz transition is clearly different, as are the column density values measured using the 1612 MHz 
transition. The discrepancy between the measured values indicates that a non-LTE excitation effect is present for OH in diffuse clouds.
The 1665 and 1667 MHz transitions should better represent the true kinematic distribution and column density of OH, and we thus exclude the 
1612 MHz profile from further discussion.  

OH is the only molecule observed in the PRIMOS data that is more typically present in the most diffuse gas of the CNM and in the WNM
than in the higher density regions. 
It is formed through warm temperature chemistry which can occur efficiently under conditions of shocks and turbulent dissipation 
\citep[see e.g.][and references therein]{Godard09, Godard14}. 
Turbulent dissipation in particular is believed to significantly impact the chemistry observed in the CNM, with 
periodic episodes of turbulent dissipation injecting significant amounts of high-temperature molecules into the CNM \citep{Godard09}. 
Under the denser, darker, cooler, and less turbulent conditions of the embedded translucent cloud material, OH should not be formed
efficiently, and the reactive radical species should be destroyed efficiently.

As expected from this, the OH absorption profiles at 1665 and 1667 MHz are more similar to those of small hydride molecules 
observed with Herschel \citep[e.g.][]{Godard12, Indriolo2015} than to other absorption profiles observed in the PRIMOS data. 
Comparing the 1667 MHz profile of OH with \cycloprop ~(Figure \ref{OH_cC3H2_profs}), it is apparent that OH is present in more diffuse material than \cycloprop.  
If \cycloprop ~traces \Htwo ~even at low extinction \citep{LisztcC3H2}, this implies that OH is present in 
material with very low values of the molecular hydrogen fraction.  
Furthermore, the OH absorption profiles indicate that OH is less abundant in the regions of enhanced density. 
This is evident as the profiles do not contain significant optical depths within the narrow peaks corresponding to 
the enhanced \cycloprop ~absorption.  These trends are consistent with the theoretical picture of OH formation.

Finally, we comment on the values of \NOH/\NHtwo ~derived.  We determine a median value of \NOH/\NHtwo ~= 1.1 $\times 10^{-6}$.
In the \los ~to  Sgr A, \citet{Karlsson13} found slightly higher values of 1.8 to 5.8 $\times 10^{-6}$ 
in \los ~clouds associated with the same structures that we observe, namely the EMR (at -130 to -95 \kms ~in our data), 
the 3-kpc arm (at -40 \kms ~towards Sgr B2), and in the 0 \kms ~gas.  The values reported here and by \citet{Karlsson13}
are significantly larger than those observed in diffuse and translucent gas towards other sightlines however, with values 
ranging from $5 \times 10^{-8}$ to $2 \times 10^{-7}$ observed towards W51 and W49N \citep{NeufeldOH, OH_otherLOS_Wiesemeyer}.  
If we consider the abundance with respect to the total hydrogen column by $N_\text{H} = N_\text{HI} + 2\,$\NHtwo, 
adopting the HI column densities published in \citet{Indriolo2015} (Table \ref{tab:NHs}), we obtain typical values of 3 to 6 $\times 10^{-7}$. 

While it is not appropriate to adopt the values of \NOH/\NHtwo ~in Table \ref{tab:abunds} as true abundances, 
as OH primarily occupies a different phase of the ISM than the \cycloprop ~from which \NHtwo ~was estimated, 
we include the values as they likely have physical significance.  
Lower than average values may indicate that a cloud is more dominated by higher density material
as opposed to WNM and CNM material.  Notably, lower than average values are observed in the -106, -40, and 0 \kms ~clouds, where 
most other molecules are particularly prominent.



\subsubsection{\formald}
\label{subsec:H2CO}

Excluding OH, \formald ~is the most abundant species measured in this work at typical abundances of a few $\times$ 10$^{-7}$.
This is significantly higher, by nearly two orders of magnitude, than measurements in high latitude diffuse clouds and cirrus clouds 
\citep{LL06, TurnerH2CO}. With respect to other interstellar environments, these abundances are very high, 
comparable to those observed in the Orion hot core and molecular ridge \citep{HEXOSsgr}, and two and four orders of magnitude, 
respectively, larger than what is observed in the dark clouds TMC-1 and in the Horsehead PDR \citep{OhishiKaifu_TMC1,Guzman14}.
In the warm kinetic temperatures and low densities of the diffuse clouds, \formald ~production 
is likely dominated by the neutral-neutral reaction of 
\begin{equation} \label{eq:H2COform} 
\text{O} + \text{CH}_3 \rightarrow \text{H}_2\text{CO} + \text{H},
\end{equation}  
which proceeds efficiently at warm kinetic temperatures of T $\sim$ ~100 K \citep{Baulch_H2CO}. 
However, under low extinction conditions, \formald ~is not predicted to be nearly as abundant as observed due to photodissociation \citep{LL06}.

The abundance of \formald  ~varies by a factor of $\sim$4 within the observed clouds and does 
not exhibit an apparent correlation with \Ncycloprop; assuming that the abundance of \cycloprop ~is constant with respect to \Htwo, 
the abundance of \formald ~therefore does not correlate with \NHtwo. However, the \formald ~abundance
does exhibit a positive linear correlation with the observed abundance of CS (Figure \ref{Abunds_CS}).  \citet{LL06} also 
demonstrated that the abundances of \formald ~and CS are tightly correlated, whereas \formald 
~is not correlated with hydrocarbons like \cycloprop.  Once again, 
the abundance ratios that we observe in the \los ~clouds to Sgr B2 are different than what has been reported in 
other sightlines;  whereas \Nformald/\NCS ~was measured to be approximately 2 to 3 by \citet{LL06}, we obtain higher values 
of $\sim$20 to 30.  Inspecting the profiles of \formald ~and CS, it is unclear why the two species should correlate well.  
Whereas CS appears to prefer higher density regions of the diffuse clouds, the absorption profile of \formald 
~includes moderately broad absorption components indicating that \formald ~is likely abundant in a larger fraction of 
the cloud volume. If a warm temperature gas phase reaction with neutral oxygen dominates the formation of \formald ~(Equation \ref{eq:H2COform}), 
then it is reasonable that \formald ~is fairly widespread.  It is curious however that the abundances of \formald ~and CS 
are correlated when they appear to occupy distinct pockets of gas. 

 
 
\subsubsection{SiO}
\label{subsec:SiO}

Abundances of SiO in the \los ~to Sgr B2(M) have been measured by \citet{GN96}, and abundance patterns in the sightlines
to Sgr B2(M) and W49\,N were discussed by \citet{GN_SiO}. In the \los ~towards Sgr B2(M), values measured by \citet{GN96} varied 
by an order of magnitude, from $3 \times 10^{-10}$ in the -40 \kms ~cloud, to 2 and $3 \times 10^{-9}$ in the -106 and 
-75 \kms ~clouds, respectively.  
In our data, we observe values in a similar range, of $1.5 \times 10^{-10}$ to $1.7 \times 10^{-9}$.  
We observe lower values in clouds located in the Galactic disk, namely $\sim4 \times 10^{-10}$ 
in the -40 and -23 \kms ~clouds, and higher values of 1.3 to 1.7 $\times 10^{-9}$ in the -73, 0, and +25 \kms ~clouds 
located in the Galactic Bar and Center.  We observe a very low value in the -58 \kms ~cloud, which is surprising 
because this is believed to be located internal to the Galactic Bar or Center.  

The profile of SiO may also hold clues to the nature of the gas in which SiO is most abundant.  The absorbing 
components at -73, -40, and -23 \kms ~in particular appear to be broader and more smoothly varying than in profiles 
of most other molecules.  The profile does not contain the narrow absorption features present in the profiles of most 
other species observed in this study, and particularly of CS and SO.
Therefore, it does not appear that SiO is enhanced in the densest material. 
The components of SiO may be shock broadened and SiO may be abundant only in recently shocked material. 
In this scenario, SiO could be confined to a small spatial region on a shock front, or could be spatially extended, 
perhaps occupying gas with more recent episodes of turbulence, which is believed to be important in the diffuse ISM
\citep{Godard09}.  Alternatively, the SiO abundance could be governed more directly by the X-ray or CR flux in the medium, 
and the species may inhabit more diffuse conditions than most molecules observed in this study.  

A deeper investigation of the formation and destruction of SiO in diffuse clouds could help elucidate the structure of 
the diffuse clouds.  For instance, if the SiO abundance is most directly governed by shocks in these environments, 
then this data may indicate that the high density clumps in which CS and SO are abundant are not subjected to 
the shocks associated with turbulent dissipation, and tend to be undisrupted within a turbulent medium.  
This would put a limit on the size scales of the clumps and advance our understanding of the physical processes ongoing in 
this material.

\subsection{Linear Hydrocarbons:  \lneutralBrett ~and \lBrett}

The linear hydrocarbon species \lneutralBrett ~has been a known interstellar molecule for more than three decades, and 
is present in multiple distinct environments including 
evolved stars \citep{lC3H_Thaddeus, Pardo07}, dark clouds \citep{lC3H_Thaddeus}, 
and PDRs \citep{Pety2012}.  The cation of this species, \lBrett, was first detected much more recently, by 
\citet{Pety2012} in the Horsehead PDR.  Whereas most molecules observed in the PDR, including \lneutralBrett 
~\citep{Pety2012}, are also detected in a dense, UV-shielded core in the Horsehead nebula \citep[e.g.][]{Guzman14}, 
\lBrett ~was not observed towards the core, providing the first indication that it is abundant only in the presence of 
far-UV radiation.  
Since the initial detection, \lBrett ~has also been detected in the Orion Bar PDR \citep{McGuire2014} 
and in material in Sgr B2 \citep{McGuire2013} that likely contains a high UV and X-ray flux \citep{Corby15, Envelope}.  
Additionally, \lBrett ~has been rigorously detected in three other PDR sources, and it is tentatively detected in two additional PDRs (B. McGuire 2015, private communication).
Despite a search for \lBrett ~towards 35 additional sources including hot cores, hot corinos, and evolved stars, 
the species was not detected \citep{McGuire2014a, McGuire2014}.
Thus, \lBrett ~appears to be an excellent indicator of a UV-enhanced environment.

Although the profile of \lneutralBrett ~is difficult to interpret due to the presence of hyperfine structure, line 
confusion, and weak signal, we determine reasonably consistent \lneutralBrett ~abundances of 1.5 to 5 $\times 10^{-10}$.
These values are slightly higher than the observed values in the Horsehead PDR of $(1.4 \pm 0.7) \times 10^{-10}$.  
Adopting the \lneutralBrett ~column derived in Sgr B2(N) \citep{McGuire2013} and the
molecular hydrogen column derived in Appendix A of \citet{Donghui_2016}, a similar value 
of 1 to 2 $\times ~10^{-10}$ is present in the PDR/XDR material in Sgr B2(N).

Measured \lBrett ~abundances in this work range from 4 to 11 $\times 10^{-11}$, although they have considerable 
uncertainties of about 50 percent given the weak signal and the presence of line blending.  
These values are slightly higher
than in the Horsehead PDR ($3 \times 10^{-11}$) and in Sgr B2 (2 to $3 \times 10^{-11}$).  For clouds with 
measured abundances of both \lneutralBrett ~and \lBrett ~in this survey, the ratio of \lneutralBrett/\lBrett 
~ranges from $\sim$2.5 to 5.5.  The ratios are therefore quite similar to the value of $\sim$4 observed in the Horsehead PDR, 
and comparable to the value of $\sim$6 observed in the PDR/XDR gas in Sgr B2.  We also note that the 
absorbing components of \lBrett ~are broader than 
observed in the profiles of most other species in this study.  While the weak signal may contribute to this, 
it may be that \lBrett ~is more spatially extended than most species observed here, preferentially populating 
the more diffuse material rather than the embedded clumps.  This would be expected if \lBrett ~is formed only in the 
presence of a strong far-UV field.

PDR models do a poor job of accounting for the abundances of most molecules observed in the diffuse 
ISM \citep[e.g.][]{Godard09, Godard14} despite the high UV fields in these environments. 
Yet it will be interesting to determine whether PDR models can account for the 
observed abundances of these two linear hydrocarbon species in the environments studied here.  
\citet{Pety2012} was able to reproduce the observed abundances of both species in the Horsehead PDR using a 
PDR model with conditions appropriate in that source.  While the mean densities are much lower in the 
diffuse and translucent clouds considered here, it is plausible that the observed abundances of 
\lneutralBrett ~and \lBrett ~may also be consistent with PDR model predictions.  A more recent study of the molecules 
suggests that they are formed from the photo-destruction of PAHs, with \lBrett ~being an intermediate to the 
formation of \lneutralBrett ~\citep{Guzman_lC3H}.  Due to the widespread presence of the Diffuse Interstellar Bands (DIBs) in 
diffuse cloud spectra \citep{LowPAHs,TielensPAHs}, it is generally accepted that PAHs are prevalent in the diffuse ISM, although 
there are some challenges to this model \citep[e.g.][]{PAHchallenges}.  Assuming that PAHs are common in the diffuse ISM, if \lneutralBrett ~and 
\lBrett ~form from the destruction of PAHs, then their abundances should not be enhanced in the higher density clumps, and
should be sensitive to PAH abundances and the UV field within the cloud.  The moderately broad absorption components that characterize the 
\lneutralBrett ~and \lBrett ~profiles are consistent with these expectations.  Further observational study of \lBrett ~and PAHs, including 
follow-up on sightlines with strong DIB features to determine whether \lBrett ~is present, could further test this model.


\section{Conclusions}
\label{sec:SACconclusions}

We have compiled the line profiles of 26 molecular lines from eleven molecules with absorption by diffuse and translucent clouds in the \los ~to Sgr B2(N) 
with the GBT PRIMOS data.  The data reveal the presence of $\sim$10 kinematically separate clouds.  In each cloud, we determined the 
column densities of observed molecules and estimated the molecular hydrogen column density by assuming a constant abundance of \cycloprop 
~with respect to \Htwo.
Conversion from the \cycloprop ~column density provides a new method for measuring the molecular hydrogen column (\NHtwo), which is, for numerous reasons described
in \S \ref{subsec:NH2_cC3H2}, 
preferable to conversion from isotopologues of more abundant molecules like CO and \HCO.  We then converted the molecular column densities 
to abundances and considered the abundance patterns.  We summarize new results below, including systematic trends with Galactocentric distance.

By comparing the \Htwo ~column densities measured by conversion from \Ncycloprop ~to previous measurements of \HI absorption \citep{Indriolo2015}, we note a stark decrease in the molecular hydrogen fraction with 
Galactocentric distance.   In the Galactic disk, we estimate a molecular fraction of 0.65, 
slightly higher than typical values of 0.4 measured in diffuse clouds that contain heavy molecules \citep{LisztcC3H2}; in material near the 3~kpc arm, the molecular 
fraction is significantly greater, at $\sim$0.85, and in the Galactic Center material, the data indicate that $\sim95$ percent of the hydrogen is in molecular form.

 We see additional trends with Galactocentric distance in the isotope ratios of \twelveThirteen, \thirtyTwoThirtyFour, and \twentyEightTwentyNine.  Whereas the 
 ratios of \formald/\formaldIso ~are higher than previous estimates of \twelveThirteen ~indicating the presence of isotope fractionation, 
 \cycloprop/\cyclopropIso ~column density ratios show a trend of increasing value with Galactocentric distance, consistent with agreed upon
 values in the Galaxy. \twentyEightTwentyNine ~ratios also exhibit a 
 gradient with Galactocentric distance, from a value of 13 in the Galactic Center to $\sim$21 in the Galactic Bar.
 On the other hand, \thirtyTwoThirtyFour, measured by CS/C$^{34}$S, does not show any apparent trend with Galactocentric distance.  
 Furthermore, values measured in this work are lower than previous measurements by a factor of $\sim$3 to 4, 
 with typical CS/C$^{34}$S of 5 to 10.
 
 The abundances of OH are significantly higher, by 1-2 orders of magnitude, in clouds along this sightline 
 compared to diffuse clouds at high Galactic longitude.  The line profiles indicate that OH preferentially 
 resides in low density gas with low molecular hydrogen fractions, in a distinct environment from the other species 
 presented in this study.  In clouds with more pronounced absorption by the other molecules featured in this paper, 
 the values of \NOH/\NHtwo ~are lower than average, so that \NOH/\NHtwo ~may indicate the degree to which a cloud is 
 dominated by very diffuse or more moderate density gas. These results are expected from our current understanding 
 of OH formation and destruction in the diffuse ISM. 
  
 Additionally, we observe linear hydrocarbon species \lneutralBrett ~and \lBrett ~in similar abundance ratios to observations in the 
 Horsehead PDR and in PDR/XDR material in Sgr B2(N).  \lBrett ~is an interesting species in that it may uniquely trace UV-enhancement, and  
 follow-up studies on the UV-induced formation of \lBrett ~and \lneutralBrett ~in diffuse clouds could prove interesting.

 The abundance trends and line profiles of CS indicate that the species may provide a good diagnostic of the physical 
 conditions and structure of diffuse and translucent clouds, and the observations indicate systematic differences in the 
 physical structure of clouds located internal to the Galactic Bar compared to the disk.  
 Whereas \cycloprop ~has been shown to have a 
 constant abundance in diffuse and translucent clouds \citep{LisztcC3H2}, the abundance of CS varies quite 
 significantly in different clouds.  No systematic offset is apparent in the abundances
 within clouds located in the Galactic Center/Bar compared to the disk, and we do not observe a trend with \NHtwo.
 However, the CS line profile includes multiple strong, narrow (\delv~\lsim~1) absorption features in clouds located 
 in the Galactic disk, suggesting the presence of multiple embedded sources with elevated CS abundances and 
 presumably higher densities.  Clouds in the Galactic Center and Bar instead contain 
 CS ubiquitously over the same moderately broad velocity ranges as \cycloprop ~and \formald.
 
 Abundances of \formald ~and SO are positively correlated with the CS abundance, confirming the correlation between \formald ~and CS 
 observed by \citet{LL06}.  Additionally CS-bearing molecules including \HCS, \HtwoCS, and CCS are many orders of magnitude overabundant compared 
 to model predictions.  These CS-bearing may prove excellent diagnostics for the cloud physical conditions, however an improved theoretical 
 treatment of sulfur-bearing chemistry is first required.
 
 Finally, the abundance of SiO varies significantly in the clouds, and the broad profiles of SiO may suggest that it is more spatially extended than most 
 of the molecules and does not reside in higher density embedded clumps.  If SiO indicates shocks in these environments, then it is apparent that the 
 -106, -73, and \zero \kms ~material is highly shocked, with the highest SiO abundance in the -73 \kms ~cloud, and that the -40 \kms ~cloud in the Galactic disk
 has recently experienced shocks. Further, if SiO inhabits recently shocked material, then the broad line profiles of SiO compared to the narrow, jagged 
 profiles of CS and SO, suggest that the dense clumps containing enhanced abundances of CS and SO are 
 not disrupted by shocks, putting constraints on their size scales.

\begin{acknowledgements}  
 J.F.C. gratefully acknowledges partial support from funds provided by the University of Virginia. 
 B.A.M. is a Jansky Fellow of the National Radio Astronomy Observatory.
 E.H. wishes to thank the National Science Foundation (US) for continued support of his research program in astrochemistry. 
 Support for this work was provided by the NSF through the Grote Reber Fellowship Program 
 administered by Associated Universities, Inc./National Radio Astronomy Observatory.
 The National Radio Astronomy Observatory is a facility of the National Science
 Foundation operated under cooperative agreement by Associated Universities, Inc. 
 \end{acknowledgements}

\bibliographystyle{aa} 
\bibliography{thesisBIB}
\label{references}

\newpage
\appendix

\section{Appendix A: Spectral Line Profiles and Computed Column Densities}
\label{sec:SAC_AppA}

In this appendix we present the line profiles of all molecular lines listed in Table \ref{tab:transitions}.  
These profiles constitute all transitions with detected \los ~absorption by the 11 molecules and 
isotopologues analyzed in this study.  We additionally provide the best fit molecular column densities derived with an assumed 
excitation temperature of \Tex = 3K, and the errors provided are determined as described in \S \ref{subsec:colDenssAbunds}.

\begin{figure*}[!htb]
\centering
\includegraphics[width=15cm]{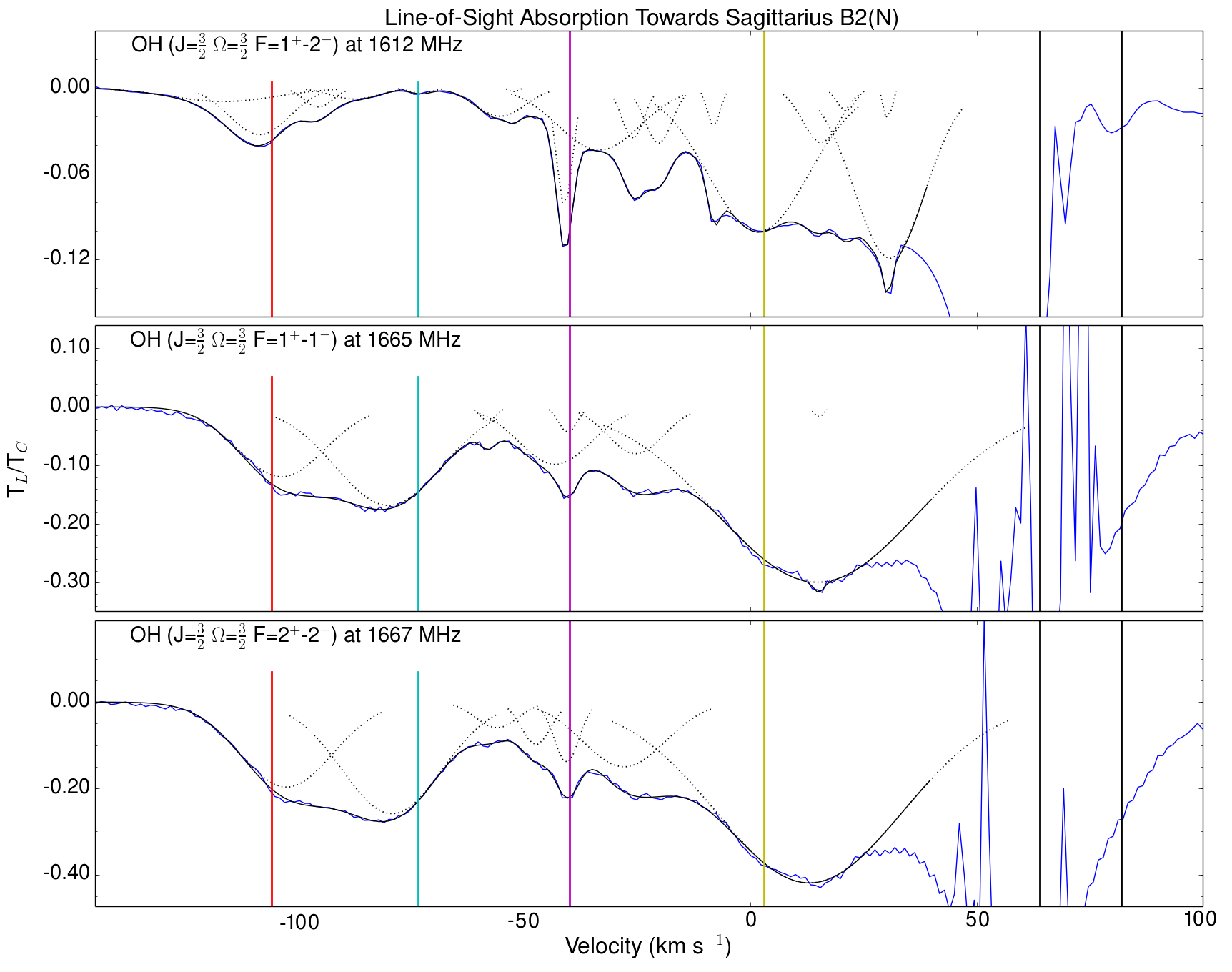}
\caption{Absorption profiles of  OH ~are shown in blue.  
Black vertical lines indicate the velocities of line absorption by Sgr B2 at +64 and +82 \kmss, and colored lines are located at -106,
-73.5, -40, and +3~\kmss.  Hyperfine or A/E structure for each of these velocity components is indicated by dotted vertical lines 
of the same color.
The data are overlaid by individual Gaussian components fit (black dotted lines), and by the total 
fit to the profile (black solid line).}
\label{ALL_OH}
\end{figure*}

\begin{figure*}
\centering
\includegraphics[width=15cm]{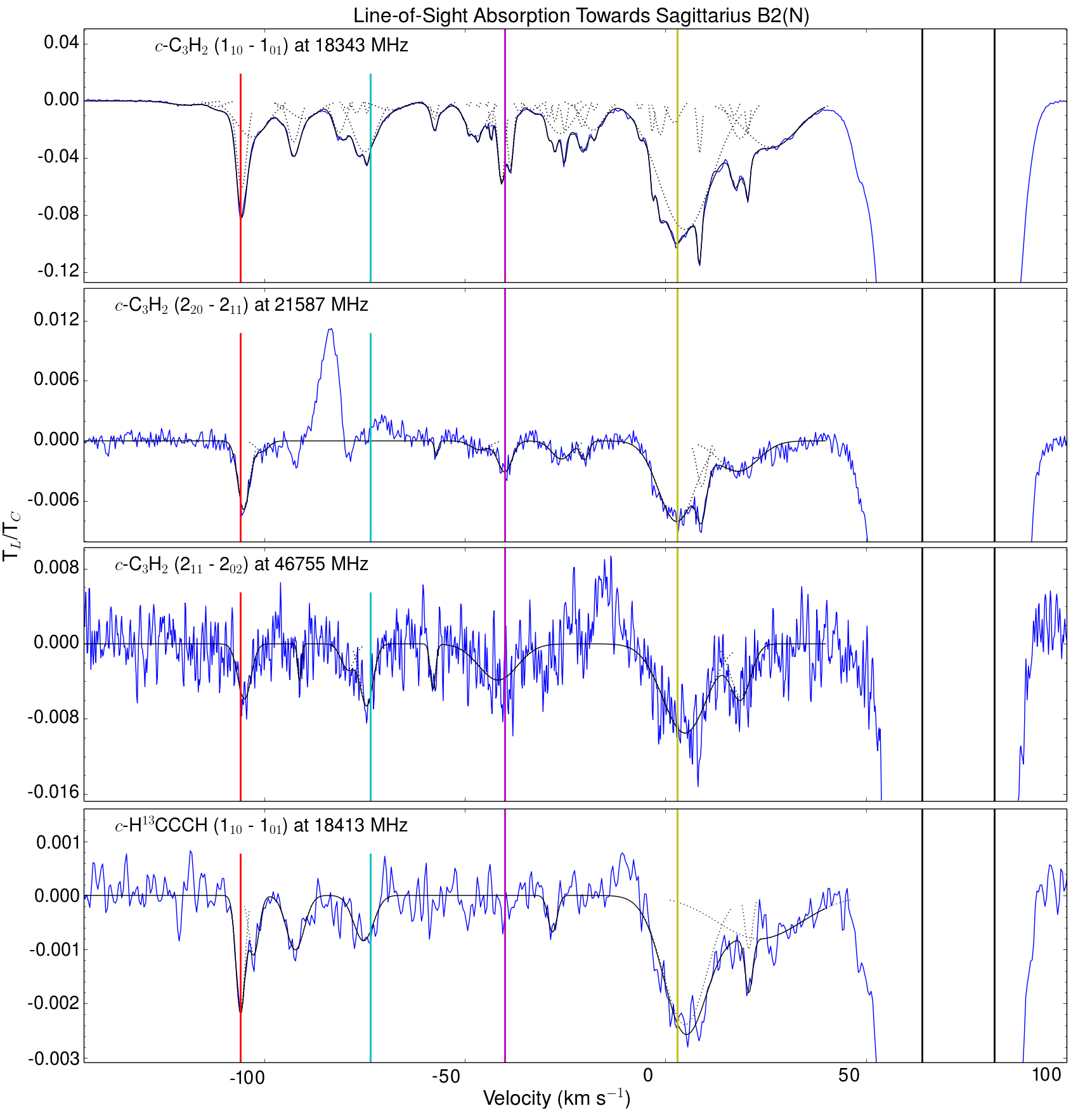}
\caption{Absorption profiles of  \cycloprop ~and \cyclopropIso ~are shown in blue.  
Black vertical lines indicate the velocities of line absorption by Sgr B2 at +64 and +82 \kmss, and colored lines are located at -106,
-73.5, -40, and +3~\kmss.  
The data are overlaid by individual Gaussian components fit (black dotted lines), and by the total 
fit to the profile (black solid line).}
\label{ALL_cC3H2}
\end{figure*}

\begin{figure*}
\centering
\includegraphics[width=15cm]{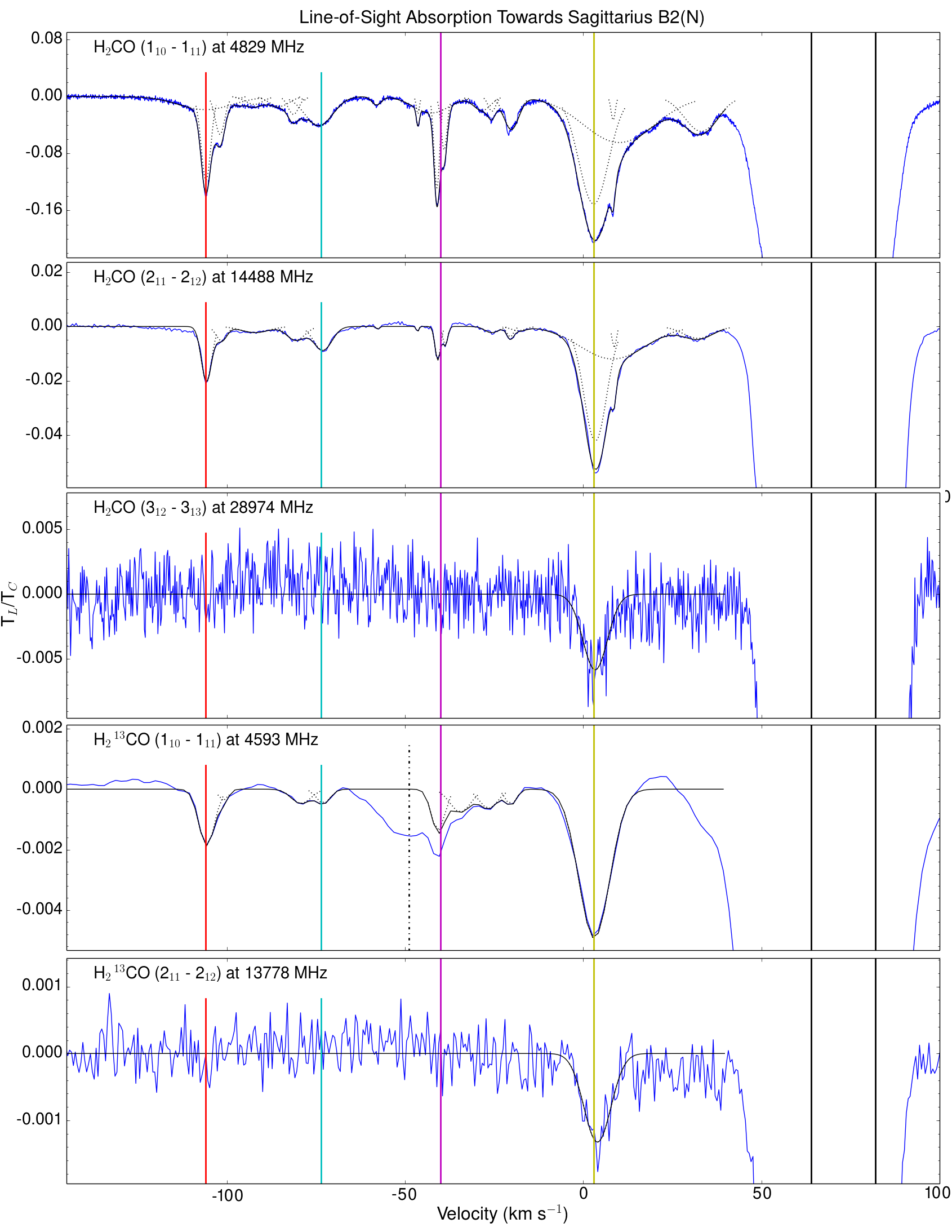}
\caption{Absorption profiles of  \formald ~and \formaldIso ~are shown in blue.  
Black vertical lines indicate the velocities of line absorption by Sgr B2 at +64 and +82 \kmss, and colored lines are located at -106,
-73.5, -40, and +3~\kmss.  
The data are overlaid by individual Gaussian components fit (black dotted lines), and by the total 
fit to the profile (black solid line).  In the profile of \formaldIso ~at 4593 MHz, the black dashed-dotted line marks an unidentified transition, and 
the black line trace of the total fit includes only absorption attributed to \formaldIso.}
\label{ALL_H2CO}
\end{figure*}

\begin{figure*}
\centering
\includegraphics[width=15cm]{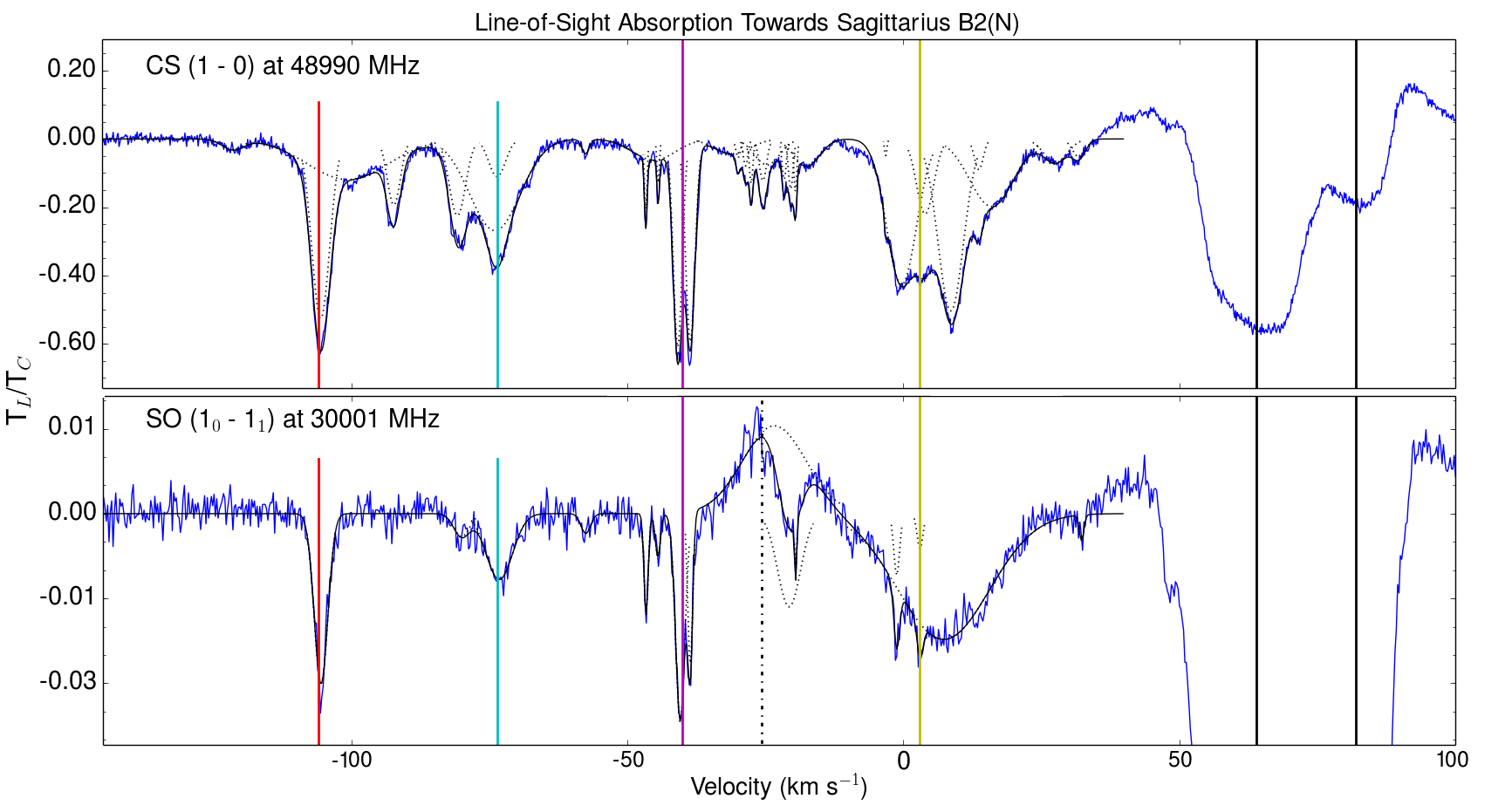}
\caption{Absorption profiles of SO and CS ~are shown in blue.  
Black vertical lines indicate the velocities of line absorption by Sgr B2 at +64 and +82 \kmss, and colored lines are located at -106,
-73.5, -40, and +3~\kmss. 
The data are overlaid by individual Gaussian components fit (black dotted lines), and by the total 
fit to the profile (black solid line).}
\label{ALL_SOCS}
\end{figure*}

\begin{figure*}
\centering
\includegraphics[width=15cm]{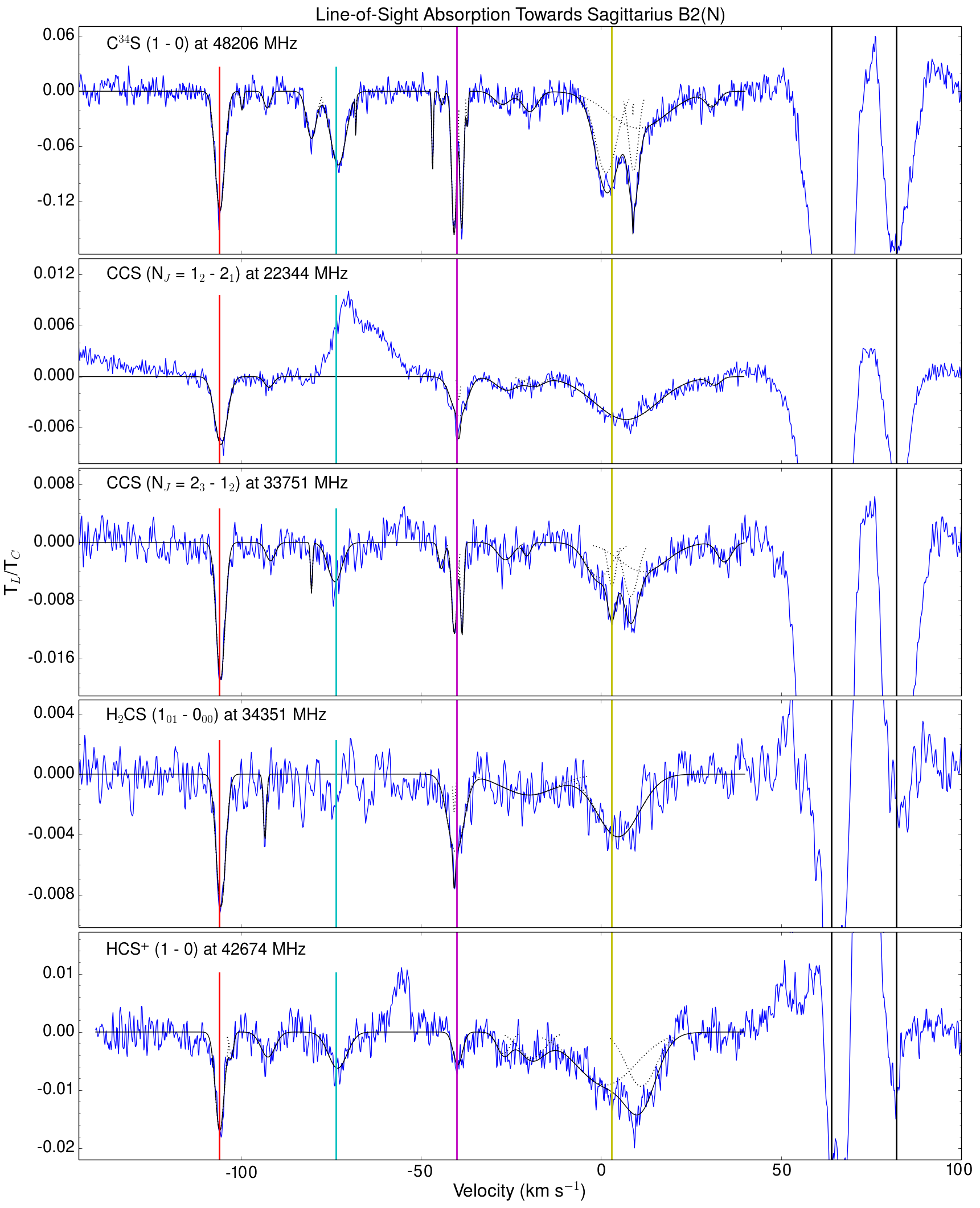}
\caption{The absorption profiles of CS-bearing species are overlaid by the best fit Gaussian components (black dotted line) 
and the sum of best-fit Gaussians (black solid line).  
Black vertical lines indicate the velocities of line absorption by Sgr B2 at +64 and +82 \kmss, and colored lines are located at -106,
-73.5, -40, and +3~\kmss. }
\label{ALL_CS}
\end{figure*}

\begin{figure*}
\centering
\includegraphics[width=15cm]{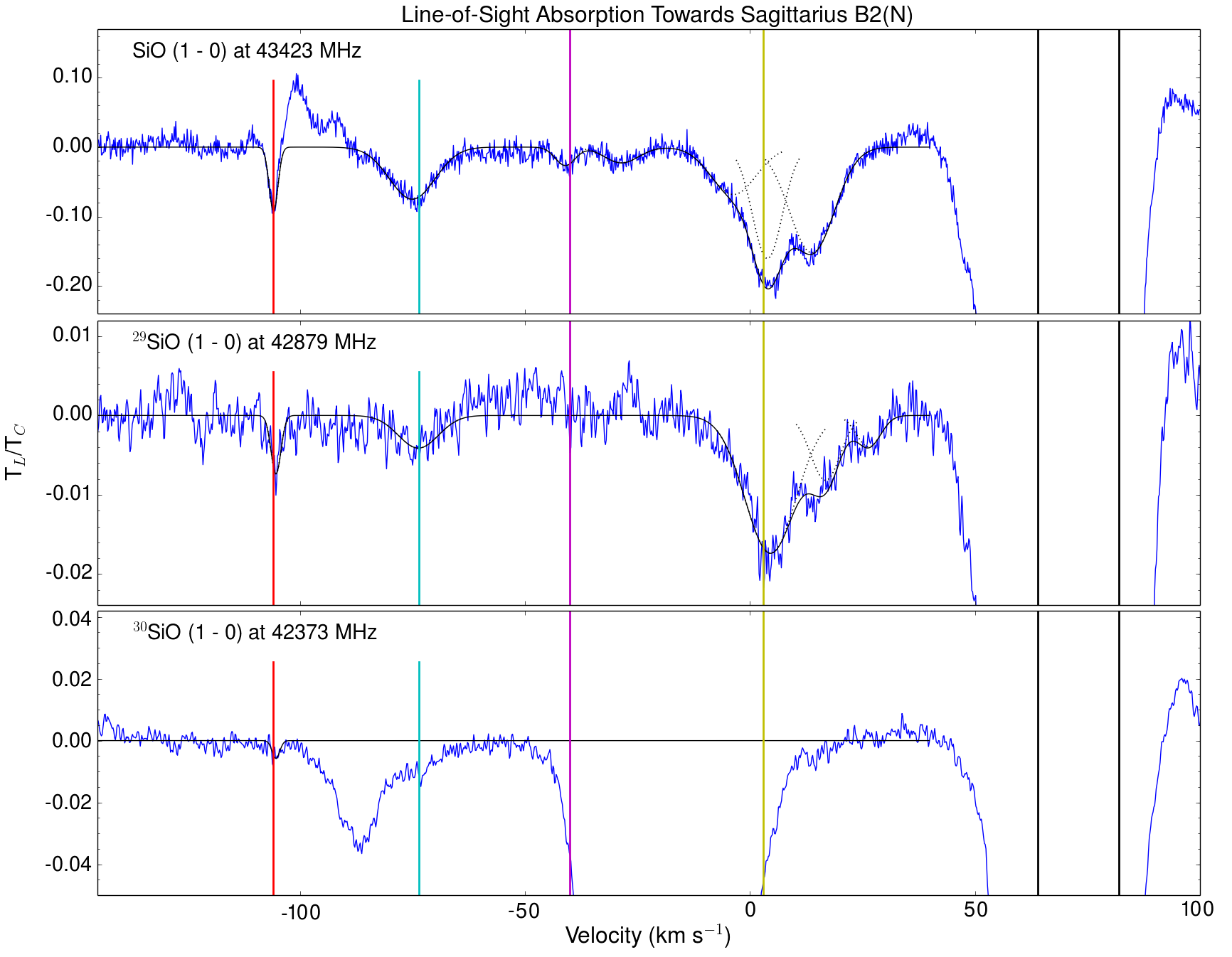}
\caption{The absorption profiles of SiO isotopologues are overlaid by the best fit Gaussian components (black dotted line) 
and the sum of best-fit Gaussians (black solid line).  
Black vertical lines indicate the velocities of line absorption by Sgr B2 at +64 and +82 \kmss, and colored lines are located at -106,
-73.5, -40, and +3~\kmss.  The profile of $^{30}$SiO includes a single detected component at -106 \kms ~due to blending with the 
\los ~absorption profile of \forma.}
\label{ALL_SiO}
\end{figure*}

\begin{figure*}
\centering
\includegraphics[width=15cm]{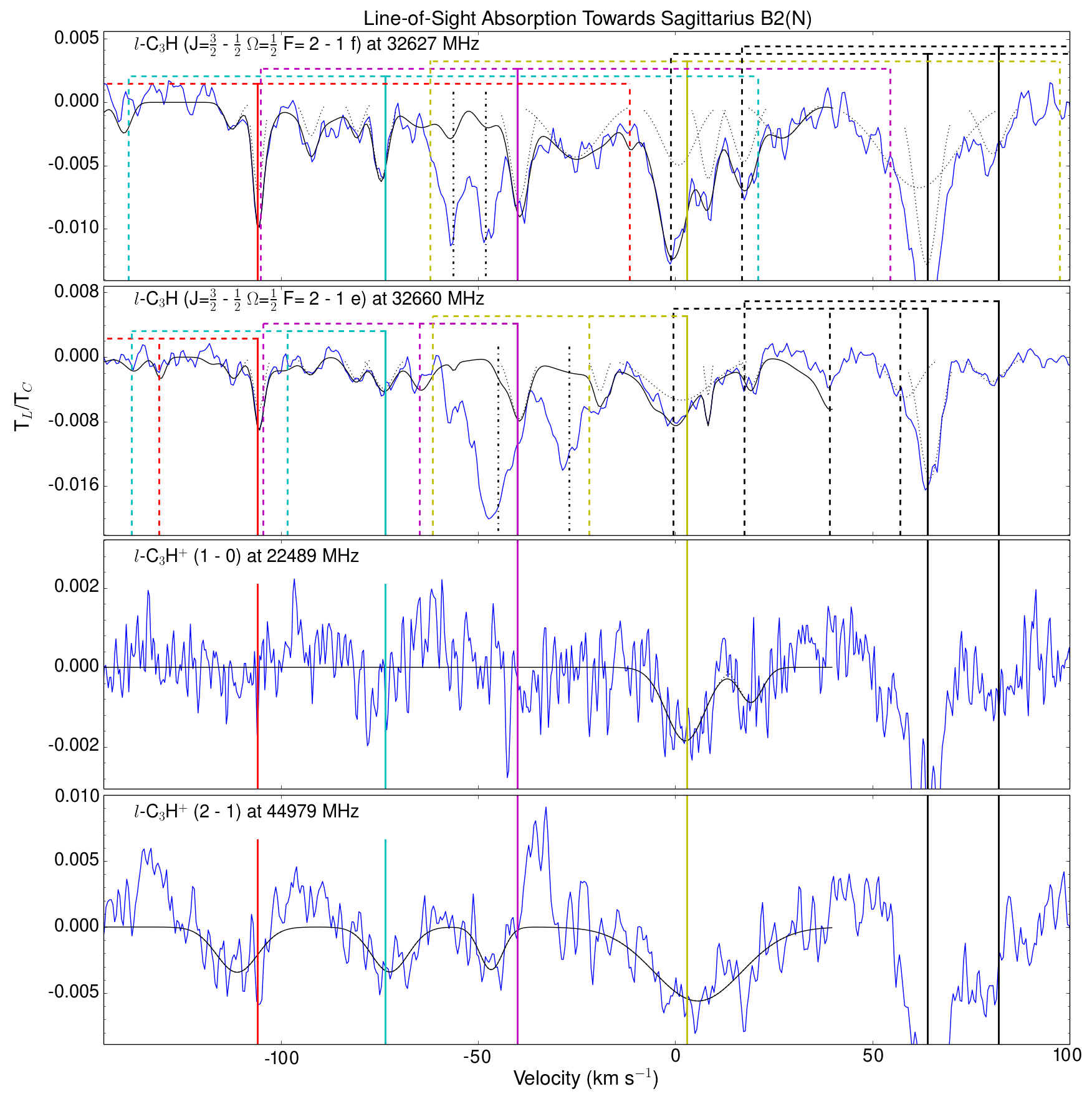}
\caption{Absorption profiles of \lneutralBrett ~and \lBrett ~are overlaid by the best fit Gaussian components (black dotted line) 
and the sum of best-fit Gaussians (black solid line).  
Black vertical lines indicate the velocities of line absorption by Sgr B2 at +64 and +82 \kmss, and colored lines are located at -106,
-73.5, -40, and +3~\kmss.  Hyperfine structure for each of these velocity components is indicated by dotted vertical lines of the same color.
The data are overlaid by Gaussian components fit to the main hyperfine component shown in black dotted lines, and by the total 
fit to the profile, which assumes that hyperfine or A/E components are present with the same line shape as the primary component, but 
with the height scaled by the ratio of the line strengths.
In the profile of \lneutralBrett ~at 32627 MHz, the black dashed-dotted line marks unidentified transitions that are not consistent with the 
typical profile of diffuse cloud absorption in this line of sight.  In the line profile of \lneutralBrett ~at 32660 MHz, the 
black dashed-dotted line marks a transition of \cisC ~at velocities of +64 and +82 \kms ~associated with Sgr B2. }
\label{ALL_lC3H}
\end{figure*}

\begin{table*}
\centering
\small
\footnotesize{
\caption{Derived molecular column densities}
\label{tab:abunds}
\begin{tabular}{lrrrrrrrrrrrl}
\hline
Species    		& \multicolumn{11}{c}{Absorption Cloud Velocity (\kmss)} 		& Units \\
			& -120	&-106	&-92	& -80	&-73	&-58	&-47	&-40	&-23	& 0 	& +20 & (cm$^{-2}$)\\
\hline
OH			&0.31(2)&2.2(1)	&2.2(1)	&1.79(6)&2.3(1)	&0.66(6)&1.1(1)	&1.2(1)	& 3.8(3)&8.4(7)	&3.4(4)	&$10^{15}$\\
\cycloprop		&0.43(4)&8.0(6)	&4.3(2)	&2.8(1)	&5.8(3)	&1.29(8)&3.5(3)	&4.5(4)	&8.8(9)	&30(3)	&7.9(13)&$10^{12}$ \\
\multicolumn{2}{l}{{\it c-}H$^{13}$CCCH}
				&3.1(7)	&1.8(6)	&	&1.4(7)	&	&  &\textless0.5&0.6(3)	&11(2)	&3.2(6) &$10^{11}$ \\
\formald 		&0.69(6)&15(3)&\textless3.1
						&4.6(5)	&7.9(3)	&1.0(1)	& 4.0(2)&12.5(6)&10.4(6)&71(20)	&8.1(8)	&$10^{14}$ \\
H$_2^{13}$CO 		&	&2.7(2)	&	&0.5(1)	& 0.8(3)&  	&	& 1.8(8)&2.2(3)	&12.6(9)& 	&$10^{13}$ \\
SO 			& 	& 3.9(3)&\textless0.4
						&0.6(1)&2.9(3)	&0.3(1)	&1.1(2)	&4.5(3)	&4.0(11)&17(1)	&3.0(2)	&$10^{12}$ \\
CS 			&0.22(2)& 5.2(3)*
					&1.54(8)&1.9(1)	&3.9(2)	&0.11(2)&0.82(5)& 4.4(2)*
											&2.5(1) &11.3(6)* 
													&1.48(7)&$10^{13}$ \\
C$^{34}$S 		&	&6.0(4)	&0.5(3)	&2.2(3)&5.9(4)&\textless0.5
									& 0.9(2)&6.5(4)&2.6(4)&20(1) 	&3.8(5)	&$10^{12}$ \\
CCS 			& 	& 8.9(6)&1.1(4)	&0.9(5)	&4.3(10)&  &\gsim1.2	& 6.2(4)&4.0(10)&22(1)	&7.9(12)&$10^{11}$ \\
HCS$^{+}$ 		& 	& 8(1)	&3(1)	&	&5.6(13)&  	&  	&2.5(7)	&\lsim9.7&36(3)	&	&$10^{11}$ \\
H$_2$CS 		& 	&11(2)	&1.7(14)&	&\textless3.1
								&  	&   	&9.9(11)&\lsim8	&23(4)&\lsim0.9&$10^{11}$ \\
SiO 			&    &$\geq$1.3&  &1.4(1)$^\dagger$&3.9(2)&\textless0.2
									&0.21(9)$^\dagger$
										&0.66(7)& 1.4(1)&20(1)	&4.2(2)	&$10^{12}$ \\ 
$^{29}$SiO 		& 	& 1.1(4)&	&0.4(3)$^\dagger$
							& 2.1(6)&  	&   	&   	&   	&14(1)	&3.3(4)	&$10^{11}$ \\
$^{30}$SiO 		& 	& 0.7(3)&	&	& 	&  	&   	&   	&   	&	&	&$10^{11}$ \\
\lneutralBrett		& 	&6(1)&2.6(9)	&2.8(6)	&6.7(13)&  	&	&\lsim11&16(3)	& 26(2)	& 7(2)	&$10^{11}$ \\
\lBrett			& 	& 1.4(8)&	&	& 2.6(8)&  	&1.6(6)	& 	& 	& 6(2)	&1.3(4) &$10^{11}$ \\

\hline
\multicolumn{13}{l}{\footnotesize{* Peak optical depths are \textgreater 0.7 in CS absorption.}}\\
\multicolumn{13}{l}{\footnotesize{$\dagger$ Appears to be a wing associated with the -73 or -40 \kms ~gas instead of a separate component.}}\\

\hline
\end{tabular}}
\end{table*}
\normalsize

\end{document}